\documentclass[aps,floats,twocolumn,nofootinbib,preprintnumbers]{revtex4}

\usepackage{amssymb}
\usepackage{amsmath}

\usepackage{graphicx}
\usepackage{graphics}
\usepackage{dcolumn}
\usepackage{color}
\usepackage{rotate}
\usepackage{fancyhdr} 
\usepackage{hyperref}
\usepackage{indentfirst}
\usepackage{multirow}

\usepackage{umoline}
\usepackage{ulem}

\usepackage{hyperref}
 \hypersetup{colorlinks, citecolor=red, linkcolor=blue, urlcolor=magenta}

\def\be{\begin{equation}}
\def\ee{\end{equation}}
\def\bea{\begin{eqnarray}}
\def\eea{\end{eqnarray}}
\def\ba{\begin{eqnarray}}
\def\ea{\end{eqnarray}}

\definecolor{darkred}{rgb}{.743,0,0}

\begin{document}

\title{Dwarf spheroidal galaxies as degenerate gas of free fermions}

\author{Valerie Domcke$^{a,b}$}
\email{valerie.domcke@sissa.it}

\author{Alfredo Urbano$^{a}$}
\email{alfredo.urbano@sissa.it}

\affiliation{ $^{a}$SISSA - International School for Advanced Studies, via Bonomea 256, 34136 Trieste (ITALY).\\
$^{b}$INFN - Istituto Nazionale di Fisica Nucleare, via Valerio 2, 34127 Trieste (ITALY).}

\preprint{SISSA 48/2014/FISI}

\date{\today}

\begin{abstract}
In this paper we analyze a simple scenario in which Dark Matter (DM) consists of free fermions with mass $m_f$. 
We assume that on galactic scales these fermions are capable of forming  a degenerate Fermi gas, in which 
stability against gravitational collapse is ensured by the Pauli exclusion principle. The mass density of the resulting configuration is governed by a non-relativistic Lane-Emden equation, thus leading to a universal cored profile that depends only on one free parameter
in addition to $m_f$.
After reviewing the basic formalism, we test this scenario against experimental data describing the velocity dispersion of the eight classical dwarf spheroidal galaxies of the Milky Way.
We find that, despite its extreme simplicity, the model exhibits a good fit to the data and realistic predictions for the size of DM halos 
providing that $m_f\simeq  200$ eV.  Furthermore, we show that in this setup larger galaxies correspond to the non-degenerate limit of the gas. We propose a concrete realization of this model in which DM is produced non-thermally via inflaton decay.
We show that imposing the correct relic abundance and the bound on the free-streaming length constrains the inflation model in terms of inflaton mass, its branching ratio into DM and the reheating temperature.

\end{abstract}

\maketitle

\section{Introduction}

The nature of Dark Matter (DM) 
is one of the most important problems yet unsolved in physics.
The most popular candidate for DM is a WIMP, a weakly interacting particle with a mass 
in the GeV-TeV range that freezes out from thermal equilibrium in the early Universe. As a consequence of this decoupling,
the WIMPs  cool off rapidly as the Universe expands. Therefore, in the WIMP scenario the DM  that we observe today is made of slowly-moving, non-relativistic cold particles.

 The cold DM scenario agrees astonishingly well with observations on cosmological scales \cite{Ade:2013zuv}.
However, cold DM  simulations do not match 
observations at small scales ($\sim$ kpc, the typical galactic scale). They incorrectly predict: \textit{i}) too many galactic bulges, \textit{ii})
steep density profiles in dwarf
galaxies (``cusp/core'' problem), \textit{iii}) too dense subhalos/satellites
(``too big to fail'' problem), \textit{iv}) too many subhalos/satellites (``missing satellites'' problem).
These problems drew an increasing level of attention, and the expression ``small-scale crisis of cold DM''
has been coined \cite{Moore:1999gc,AvilaReese:2000hg,Maccio:2005wr,BoylanKolchin:2011dk}.

In order to solve these problems, 
two different approaches are commonly pursued.
1) Introduction of baryons \cite{RomanoDiaz:2008wz,RomanoDiaz:2009yq}. Including models for baryons in the Universe can significantly alter
the results from structure formation simulations.
2) Alternative DM paradigm. Along this line, e.g., models of fermionic warm DM (WDM) \cite{Colin:2000dn,Bode:2000gq} and self-interacting DM \cite{Spergel:1999mh} have been studied. Recently, 
the possibility that DM is made of ultra-light bosons able to form a Bose-Einstein condensate (BEC) on galactic scales has been proposed \cite{Sin:1992bg,Ji:1994xh,Boehmer:2007um,Harko:2011jy,Chavanis:2011uv,Harko:2011xw,Harko:2011zt,Pires:2012yr,Bettoni:2013zma,Diez-Tejedor:2014naa,Li:2013nal,RindlerDaller:2011kx} (see refs.~\cite{Sikivie:2009qn,Davidson:2014hfa,Davidson:2013aba,Saikawa:2012uk} for the special case in which the BEC is made of QCD axions, and ref.~\cite{Hu:2000ke} for a more general scenario dubbed fuzzy DM). 
Neglecting for simplicity self-interactions, in a BEC the pressure supporting the system from gravitational collapse  is provided by the
Heisenberg uncertainty principle.
However, this mechanism alone is capable to sustain a galactic structure only if the boson 
is ultra-light, i.e.\ with a mass of order $\mathcal{O}(10^{-25}~{\rm eV})$.

The situation is completely different if
we consider a fermion rather than a boson,
since the pressure arising from the Heisenberg uncertainty principle is enforced by the Pauli exclusion principle.
As a consequence, galactic structures can be protect from gravitational collapse even 
for larger values of the DM mass.

In refs.~\cite{Destri:2013pt,deVega:2013woa,deVega:2013jfy} 
the observed properties of galactic structures have been studied in the context of WDM. 
The starting assumption is that a galactic halo can be described by a self-gravitating Fermi gas of DM particles.
The outcome of the analysis outlines an extremely interesting scenario. In particular, it is argued that 
compact dwarf galaxies correspond to the quantum degenerate limit
of the Fermi gas, while larger galaxies 
correspond to the classical Maxwell-Boltzmann limit.

In this paper we are interested in the degenerate Fermi limit, and our aim is to test 
this scenario against the experimental data describing the kinematic of the Milky Way's dwarf spheroidal galaxies. {Since these galaxies are 
completely dominated by DM, it is reasonable to expect that all the observed kinematic data are tightly linked to the fundamental properties of their DM content.}
Moreover, since the possibility to form a degenerate Fermi configuration only relies on the fermionic nature of DM particles, we 
are not forced to have in mind any specific WDM candidate, like for instance sterile neutrinos.
On the contrary, our analysis assumes the simplest DM candidate one can imagine: a free fermion with mass $m_f$.

In more detail, 
this paper is organized as follows. In section~\ref{sec:QuantumProperties} 
we sketch, using qualitative arguments, the role of the quantum pressure. 
In section~\ref{sec:FermiGasQ} we review the basic properties of a degenerate Fermi gas model.
In section~\ref{sec:velocity} we fit the  model against the data describing the velocity dispersion of the eight classical 
dwarf spheroidal galaxies of the Milky Way.
In section~\ref{sec:Discussion} we discuss the results
 in the context of an explicit realization of the model in which DM is produced non-thermally via inflaton decay. In this section we also test the hypothesis that spiral galaxies can be described by non-degenerate configurations of the same DM particle.
Finally, we conclude in section~\ref{sec:Conclusions}. 
In appendix~\ref{app:Fit} we collect the details of the fit performed in section~\ref{sec:velocity}.
In appendix~\ref{app:FermiDirac}
we shortly review the statistical mechanics of self-gravitating fermions.

\section{Quantum pressure versus gravity}\label{sec:QuantumProperties}

Galaxy formation is driven by classical gravitational physics.
However,  in order to protect a galaxy from gravitational collapse,
one needs to counterbalance gravity 
with a supporting pressure, 
ordinarily given by thermal pressure. 

Here, we want to explore the possibility that the equilibrium 
is sustained by quantum pressure. 
Quantum pressure arises from the two fundamental principles of quantum theory: the Heisenberg uncertainty principle
and the Pauli exclusion principle. The quantum pressure 
 is present even when the temperature of the gas is so low that the ordinary pressure does not hold it up. 
This is exactly the degenerate limit we are interested in.
For bosons, the degenerate configuration 
is the BEC; for fermions, it corresponds to the degenerate Fermi gas.
In the following, we will discuss both these situations. 
The aim of this section is to provide an intuitive 
picture 
of the physics involved.
The system we are interested in is a gas of $N$ fermionic (bosonic) DM particles with mass $m_{f}$ ($m_b$), confined in a volume $V$, 
with number density $n=N/V$, total mass $M$, radius $R$, and mass density $\rho = M/V$. 

\subsection{Ultra-light bosons and the BEC}

First, let us illustrate the bosonic case.
We start our discussion by computing the smallest momentum allowed by the Heisenberg uncertainty 
principle. The minimum momentum -- minimum since we are interested in the ground state of the system -- is associated with the 
maximum uncertainty in the position, i.e.\ $\Delta x\sim R$. Therefore, the minimum momentum is given by $p\sim h/R$, where $h$ is the Planck constant. 
The quantum pressure $P_{\rm Q}$, i.e.\ the flux of Heisenberg momentum, is given by $P_{\rm Q} \sim nvp \sim h^2\rho/m_b^2 R^2$,
where we used $n=\rho/m_b$, and the typical momentum-velocity relation for non-relativistic particles.
Remarkably, this qualitative result
can be obtained in a more formal way using the Gross-Pitaevskii-Poisson 
description of a self-gravitating BEC \cite{Boehmer:2007um,Harko:2011jy,Chavanis:2011uv,Harko:2011xw,Harko:2011zt,Pires:2012yr}. 
 The pressure from gravitational attraction, i.e.\ the gravitational force per unit of area, is given by 
$P_{\rm G}\sim GM^2/R^4$, where $G$ is the Newton constant. Therefore, the condition  for  
equilibrium is given by
\begin{equation}
{\rm Bosons:}~~~P_{\rm Q} \sim P_{\rm G}~~\Rightarrow~~\frac{h^2\rho}{m_b^2 R^2} \sim \frac{GM^2}{R^4}~.
\end{equation}
Using the dimensional estimate $\rho \sim M/R^3$ we extract
the typical size of a self-gravitating BEC with mass $M$
\begin{equation}
R \sim  \frac{h^2}{GMm_b^2}~.
\end{equation}
Considering for concreteness a typical size $R\sim 100$ kpc, $M = 10^{12}$ M$_{\odot}$ (with solar mass ${\rm M}_{\odot}=1.98\times 10^{30}$ kg), one obtains $m_b \sim 10^{-25}$ eV.

\subsection{Degenerate Fermi gas}

{Let us now move to discuss what happens considering fermions instead of bosons.
For a similar discussion, see also refs.~\cite{Destri:2013pt,Destri:2012yn}.}
According to the Pauli exclusion principle 
if there are $N$ fermions in the system 
they can not occupy the same state 
with minimum momentum -- as they would if they were bosons -- but they 
must form pairs with increasing value of momentum separated by at least $p\sim h/R$.
The minimum momentum 
must therefore be $p\sim N^{1/3}h/R$. {Therefore, if $N\gg 1$ the quantum pressure for a system of
 fermions is much bigger than in the bosonic case.} The related quantum pressure is given by
$P_{\rm Q}\sim nvp\sim h^2\rho^{5/3}/m_f^{8/3}$ where we used  $n=\rho/m_f$ and $V\sim R^3$. 
The equilibrium condition is 
\begin{equation}
{\rm Fermions:}~~~P_{\rm Q} \sim P_{\rm G}~~\Rightarrow~~\frac{h^2\rho^{5/3}}{m_f^{8/3}} \sim \frac{GM^2}{R^4}~.
\end{equation}
Using again the dimensional estimate $\rho \sim M/R^3$ we extract
the typical size of a self-gravitating degenerate Fermi gas with mass $M$
\begin{equation}
R\sim \frac{h^2}{m_f^{8/3}GM^{1/3}}~.
\end{equation}
Considering for concreteness the typical size $R\sim 100$ kpc, $M = 10^{12}$ M$_{\odot}$, one obtains $m_f \sim 20$ eV.
Therefore, a degenerate Fermi gas may sustain a galactic structure with $m_f \gg m_b$.
Motivated by these results,
in the next section we will study the details of the degenerate Fermi gas configuration.

\section{Degenerate Fermi gas: quantitative analysis}\label{sec:FermiGasQ}

The assumption underlying this paper is that in the present Universe a galactic structure  
can be described  by  a self-gravitating Fermi gas of DM particles in a state of statistical equilibrium. More formally, this equilibrium state is represented by a Fermi-Dirac distribution.
 Throughout our analysis, we neglect the role of baryons. This assumption is perfectly justified even in the case of large galaxies, where the baryonic component covers at most a few percent of the total mass \cite{Persic:1995ru}. Following refs.~\cite{Destri:2013pt,deVega:2013woa,deVega:2013jfy}, we assume the same distinction between small and large galaxies: the former are close to the degenerate limit of the Fermi-Dirac distribution, the latter are described by the opposite Maxwell-Boltzmann regime. As mentioned in the introduction, in this paper we are interested in the degenerate limit of the Fermi gas, and we aim to test it against the kinematic data describing the classical dwarf spheroidal galaxies of the Milky Way. In the following, we describe quantitatively the degenerate limit (see appendix~\ref{app:FermiDirac} for a more detailed discussion); in particular, we provide all the formulas that are relevant for the phenomenological analysis that will be performed in section~\ref{sec:velocity}. Most of the arguments faced in this section are known from standard textbooks. We refer to ref.~\cite{Padmanabhan1} for a general introduction.

\subsection{General overview}\label{sec:FermiGas}

At $T=0$ the Fermi-Dirac distribution is a step function
\begin{equation}
f_{\rm FD}=\left\{
\begin{array}{cc}
1  &~~~p\leqslant p_{\rm F}   \\
0  &~~~p > p_{\rm F}
\end{array}\right.~,
\label{eq:FDT0}
\end{equation}
where $p_{\rm F}$ is the Fermi momentum.
Since the temperature is zero, particles have zero kinetic energy. If they were bosons, they would occupy
the lowest energy level; fermions, on the contrary, are subject to the Pauli exclusion principles, and they will fill all states 
with momentum lower than $p_{\rm F}$. It is possible to associate to the Fermi momentum a Fermi velocity $v_{\rm F} = p_{\rm F}/m_f$.
The number of levels with momentum between $p$ and $p+dp$ is given by
$d\chi = (4\pi/h^3)p^2dp$; since there are two particles for each level, the number of fermions 
per unit of volume\footnote{In general the number of internal (spin) degrees of freedom is $g= 2s+1$. In this paper we focus on the case $g=2$.} is $n=\int_0^{p_{\rm F}} 2d\chi = 8\pi p_{\rm F}^3/3h^3$. The mass density is 
$\rho = mn= 8\pi m p_{\rm F}^3/3h^3$, from which we get $p_{\rm F} = (3h^3\rho/8\pi m)^{1/3}$.
Knowing $p_{\rm F}$,
we can compute the pressure via the usual integral $P=(8\pi/3h^3)
\int_0^{p_{\rm F}} (p^4/\sqrt{p^2 + m_f^2}) dp$.
In the non-relativistic case we obtain  
\begin{equation}
P = \frac{h^2}{5m_f^{8/3}}\left(\frac{3}{8\pi}\right)^{2/3} \rho^{5/3}~,
\end{equation}
while in the ultra-relativistic case
\begin{equation}
P = \frac{h}{8m_f^{4/3}}\left(\frac{3}{\pi}\right)^{1/3} \rho^{4/3}~.
\end{equation}
A given configuration of matter 
will be in equilibrium if the gradient of the pressure is balanced by the gravitational attractive force per unit of volume. More formally, 
the equation for the hydrostatic equilibrium is given by
\begin{equation}\label{eq:HE}
\frac{dP(r)}{dr} = -\frac{GM(r)}{r^2}\rho(r)~.
\end{equation}
Throughout this paper we assume spherical symmetry; 
in eq.~(\ref{eq:HE})
$M(r)$ is the mass within the radius $r$,
and the same radial dependence is
explicitly written also for the pressure and the mass density.
A simple order-of-magnitude estimate 
reveals that in the non-relativistic case $dP/dr \sim M^{5/3}/R^6$, while in the ultra-relativistic case 
$dP/dr \sim M^{4/3}/R^5$, where $M$ and $R$ are, respectively, the characteristic mass and length scale 
of the configuration. On the other hand, from eq.~(\ref{eq:HE}), the gravitational force per unit of volume scales according to the ratio
 $M^2/R^5$. In the non-relativistic case the Pauli pressure and the gravitational force depend on the
 radius with a different power; it means that -- for a given value of mass $M$ -- the gas can always adjust
 the radius until the two forces are equal. In the ultra-relativistic case the Pauli pressure and the gravitational force
 depend on the radius in the same way; therefore, equilibrium is possible only 
 for one value of the mass. As a consequence, it is important to 
 understand under which conditions either of the two limits is realized. To this purpose, it is useful 
 to define the critical density $\rho_{\rm crit}$ as the density at which the Fermi 
 momentum becomes equal to the fermion mass, i.e.\ $m_f = (3h^3\rho_{\rm crit}/8\pi m_f)^{1/3}$, from which we get
 \begin{equation}\label{eq:CV}
 \rho_{\rm crit} = 8.3\times 10^{-24}~{\rm kg}~{\rm cm}^{-3}\left(\frac{m_f}{{\rm eV}}\right)^4~.
 \end{equation}
 If $\rho \ll \rho_{\rm crit}$ ($\rho \gg \rho_{\rm crit}$), the gas is non-relativistic (ultra-relativistic) since $p_{\rm F}\ll m_f$ ($p_{\rm F} \gg m_f$).
As we shall see later, for $m_f > 1$~eV the critical value in eq.~(\ref{eq:CV}) is well above the typical 
mass density characterizing DM halos of galactic size. In the analysis of the dispersion velocities of the dwarf spheroidal galaxies, therefore, 
we will always use the non-relativistic limit. Nevertheless, it is instructive to keep both limits for the rest of this discussion, writing in full generality 
the equation of state in the form 
$P = K\rho^{\gamma}$, with 
$K \equiv h^2/5m_f^{8/3}(3/8\pi)^{2/3}$ ($K \equiv h/8m_f^{4/3}(3/\pi)^{1/3}$) and polytropic index $\gamma = 5/3$ ($\gamma = 4/3$) in 
the non-relativistic (ultra-relativistic) limit. Coupling eq.~(\ref{eq:HE}) with the continuity equation $dM/dr = 4\pi r^2 \rho(r)$, we obtain 
\begin{equation}\label{eq:Density}
\frac{1}{r^2}\frac{d}{dr}
\left(
\frac{r^2}{\rho}K\gamma \rho^{\gamma - 1}\frac{d\rho}{dr}
\right)
=
-4\pi G\rho~.
\end{equation}
Using $\gamma = 1+1/n$ -- with $n=3/2$ ($n=3$) in the non-relativistic (ultra-relativistic) limit -- and rescaling the radial coordinate according to
 $\xi = r/\alpha$, with 
 \begin{equation}
\alpha \equiv \left[
\frac{
(n+1) K \rho_0^{1/n -1}
}{
4\pi G
}
\right]^{1/2}~,
\end{equation}
it is straightforward to show that eq.~(\ref{eq:Density})
is equivalent to the Lane-Emden equation
\begin{equation}\label{eq:LE}
\frac{1}{\xi^2}\frac{d}{d\xi}\left(
\xi^2\frac{d\theta}{d\xi}
\right) = -\theta^n~,
\end{equation}
where $\theta$ is related to the density via $\rho = \rho_0\theta^n(\xi)$,
for central density $\rho_0$. The Lane-Emden equation~(\ref{eq:LE})
can be solved numerically for the values of $n$ that are relevant in the present analysis, using the boundary 
conditions $\theta(0)=1$, $\theta^{\prime}(0)=0$. The first zero of the solution, $\theta(\xi_1)=0$, defines the radius of the configuration
$R = \xi_1\alpha$ while the total mass is given by $M=4\pi\int_0^R r^2 \rho(r)dr$. We find
\begin{eqnarray}
R&=&\xi_1 \left[\frac{(n+1)K}{4\pi G}\right]^{1/2}
\rho_0^{(1-n)/2n}~,\label{eq:R}\\
M&=&
4\pi\xi_1^2
|\theta^{\prime}(\xi_1)|
\left[
\frac{
(n+1)K
}{4\pi G}
\right]^{3/2}
\rho_0^{(3-n)/2n}~.\label{eq:M}
\end{eqnarray}
Combining eq.~(\ref{eq:R}) and eq.~(\ref{eq:M}),
we find the mass-radius relation
\begin{equation}\label{eq:MRrelation}
M = 
4\pi \xi_1^{\frac{1+n}{n-1}}
|\theta^{\prime}(\xi_1)|
\left[
\frac{(n+1)K}{4\pi G}
\right]^{n/(n-1)}
R^{\frac{3-n}{1-n}}~.
\end{equation}
For definiteness, we find the numerical approximations
$\xi_1 = 3.65$, $\xi_1^2\theta^{\prime}(\xi_1)=-2.714$ for $\gamma = 5/3$ 
($\xi_1 = 6.89$, $\xi_1^2\theta^{\prime}(\xi_1)=-2.018$ for $\gamma = 4/3$).

Besides these analytical results, it is important to keep in mind -- using more qualitative arguments -- the general features of the model.
The crucial physical properties can be
understood looking again at the dimensional analysis of the equilibrium condition in eq.~(\ref{eq:HE}), that we rewrite here for 
convenience in the non-relativistic limit we are interested in
\begin{equation}
\frac{dP(r)}{dr} =-\frac{GM(r)}{r^2}\rho(r)~~\Rightarrow~~
\frac{M^{5/3}}{R^6}\sim \frac{M^2}{R^5}~.
\end{equation}
Suppose now to have an equilibrium condition for a given mass and radius $M$, $R$. If we increase the mass,
the gravitational force per unit of volume grows
faster than the repulsive force induced by the Pauli pressure. To maintain the equilibrium,
the system 
decreases its radius, since the Pauli pressure increases faster than the gravitational force going towards smaller distances. As a consequence, the mean density of the 
configuration, $\langle \rho\rangle = 3M/4\pi R^3$, increases. In a degenerate Fermi configuration, 
larger values of mass correspond to more compact objects.

\begin{figure}[!htb!]
\begin{center}
   \includegraphics[scale=0.65]{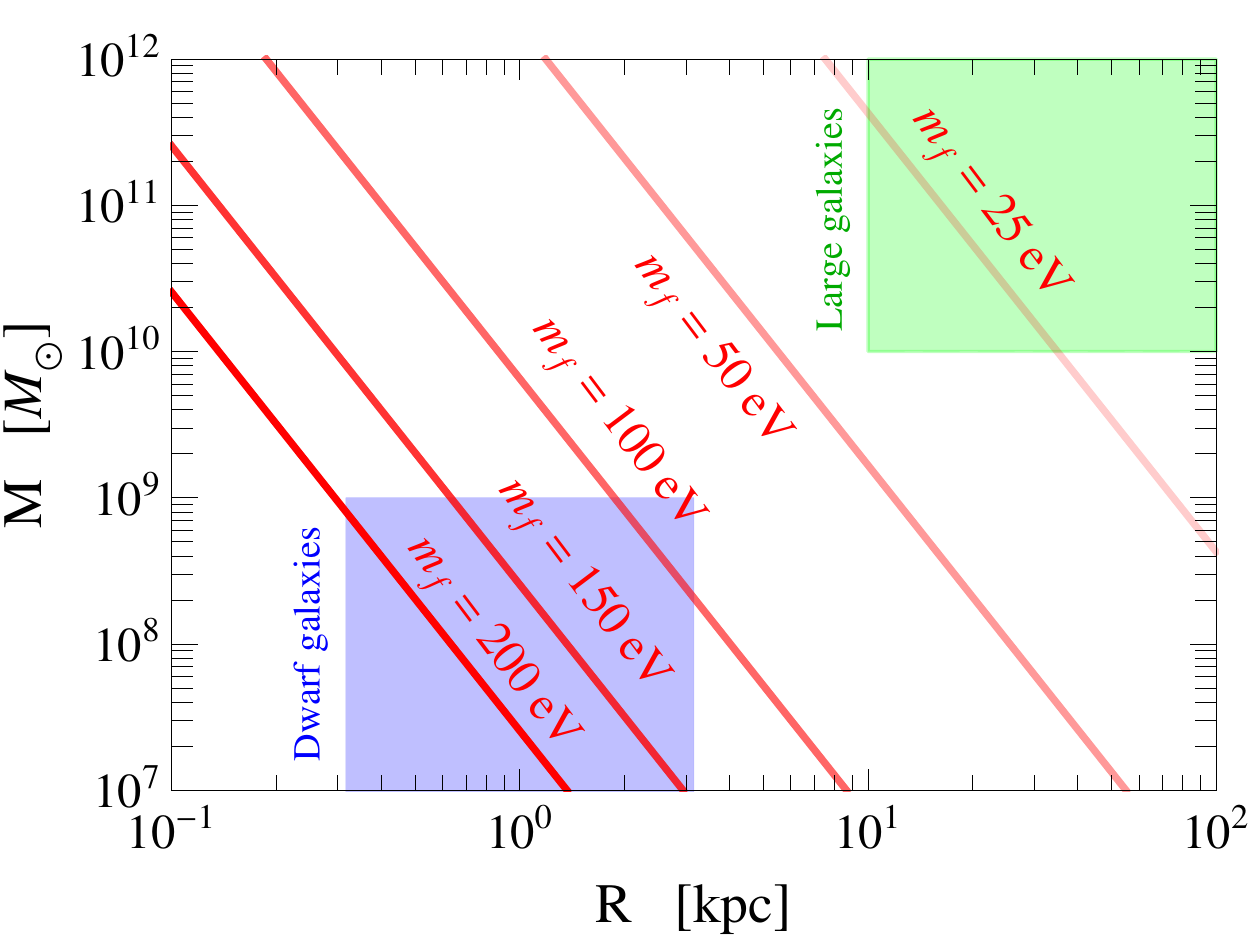}
\caption{\textit{
Mass-radius relation in eq.~(\ref{eq:MRrelation})
for a non-relativistic degenerate Fermi gas for different values of the DM mass $m_f$.
For illustrative purposes, we also show the typical size of dwarf galaxies (blue shaded)
and large galaxies (green shaded region).
}}
\label{fig:MassRadioRelation}
\end{center}
\end{figure}

\begin{figure*}[!htb!]
\begin{center}
\centering
  \begin{minipage}{0.47\textwidth}
   \centering
   \includegraphics[width = \textwidth]{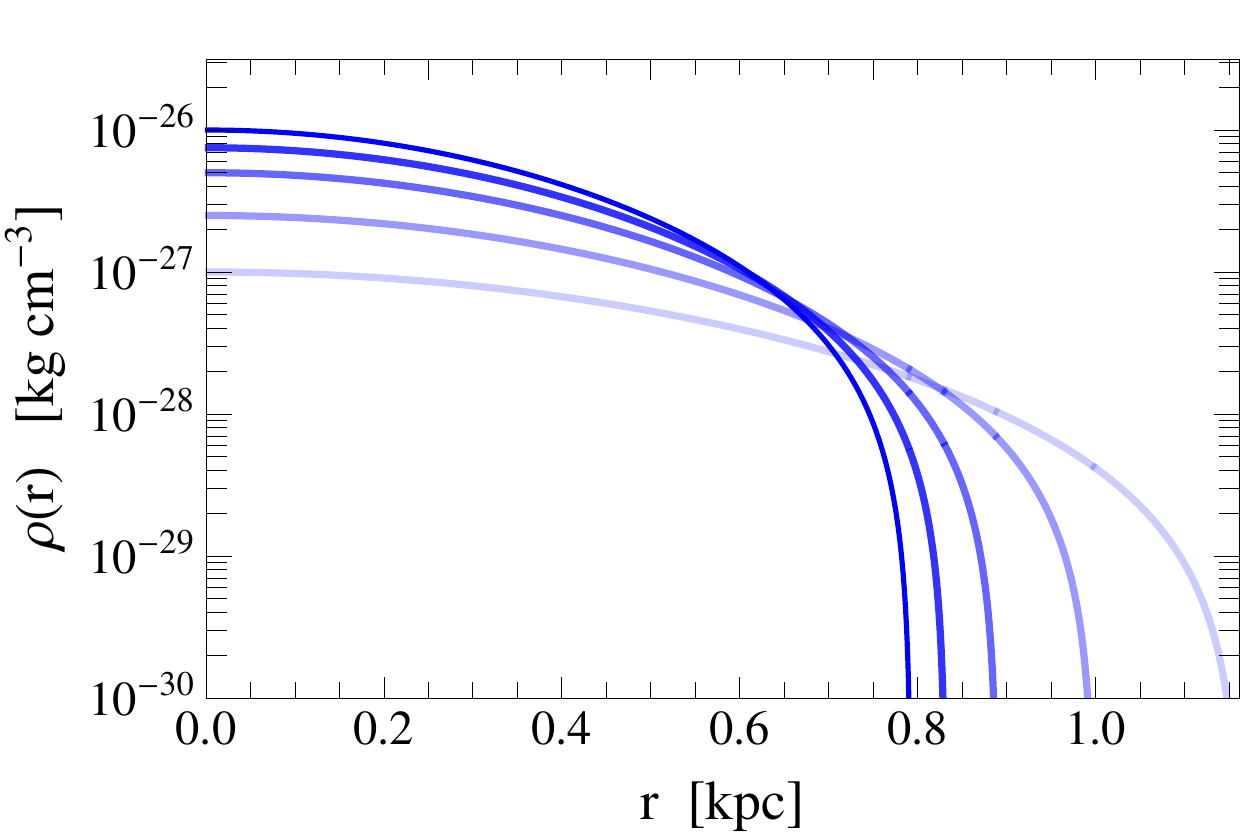}
    \end{minipage} \hfill
   \begin{minipage}{0.47\textwidth}
    \centering
    \includegraphics[width = \textwidth]{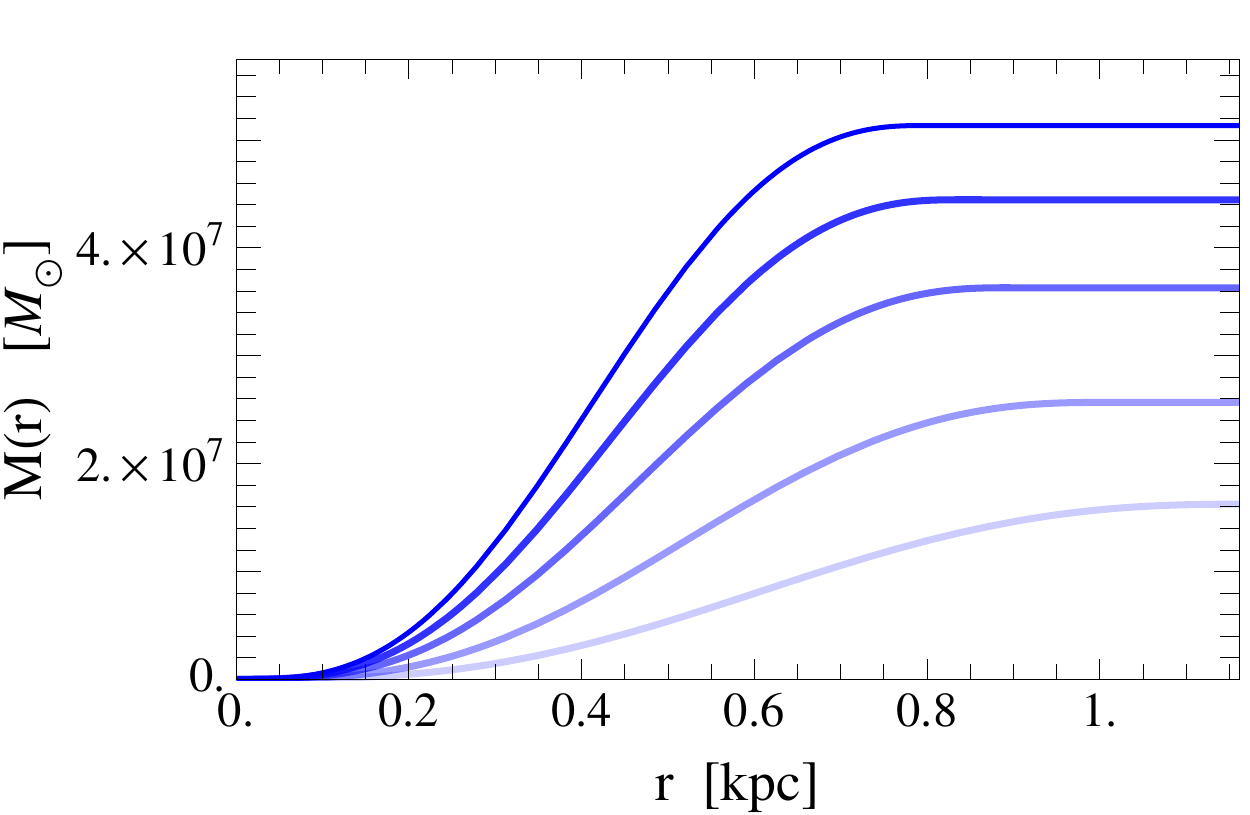}
    \end{minipage}
\caption{\textit{
Density (left panel) and mass (right panel) distributions as a function of the radius for a degenerate Fermi gas described 
by the Lane-Emden equation~(\ref{eq:LE}).
We show different values of the central density (from light to dark, 
$\rho_0 = 10^{-27},\,2.5\times 10^{-27},\,5\times 10^{-27},\,7.5\times 10^{-27},\,10^{-26}$ kg/cm$^3$). Here we take $m_f = 200$ eV.
}}
\label{fig:MassAndDensity}
\end{center}
\end{figure*}

We plot the mass-radius relation, eq.~(\ref{eq:MRrelation}), in fig.~\ref{fig:MassRadioRelation} for the non-relativistic case.
From this plot, it is clear that large galactic structures (i.e.\ galaxies with representative values of total mass and radius equal to
$M\sim 10^{10}$\,-\,$10^{12}$~M$_{\odot}$, $R\sim 10$\,-\,$100$ kpc) can be reproduced only considering small values of $m_f$, i.e.\ $m_f\lesssim 10$ eV, while,
on the contrary, dwarf galaxies  (i.e.\ galaxies with representative values of total mass and radius equal to
$M\sim 10^{7}$\,-\,$10^{9}$ M$_{\odot}$, $R\sim$ few kpc) require larger values, i.e.\ $m_f \gtrsim 100$~eV. In the remainder of this paper,
we will mainly focus on the classical dwarf spheroidal galaxies of the Milky Way, and in section~\ref{sec:velocity} we will test the degenerate Fermi gas model against experimental data
describing their velocity dispersion. Dwarf galaxies, in fact, are astrophysical objects largely dominated by their DM component - as inferred from the 
analysis of the stellar-to-halo mass ratio \cite{Mateo:1998wg,DiCintio}. The velocity
dispersion of the dwarf spheroidal galaxies remains approximately flat with radius, thus suggesting a core profile for the DM mass density 
and  a mass $M(r)$ linearly increasing with the radial distance. In the left panel of fig.~\ref{fig:MassAndDensity}, 
we show the mass density $\rho(r)$, solution of eq.~(\ref{eq:Density}), as a function of the radius for a fixed value of mass, $m_f = 200$ eV, 
and different values of the central density $\rho_0$. In the right panel  of fig.~\ref{fig:MassAndDensity}, 
we show the mass $M(r)$ of the configuration as a function of the radius. On the qualitative level, both 
the mass density and the mass profile 
seem to possess the right prerogatives to fit the observed galactic rotation curves:
the former clearly exhibits a core profile, the latter a linear increase with $r$. 
It is striking to observe how these two properties are directly connected to the physical assumptions underlying the degenerate Fermi gas model,
 instead of be the outcome of a complicated numerical simulation.
We postpone to section~\ref{sec:velocity} a careful phenomenological analysis.  

Finally, notice that for $n=3$, corresponding to the ultra-relativistic limit, the radius of the configuration, $R$, disappears from 
eq.~(\ref{eq:MRrelation}) thus defining -- for a fixed value of $m_f$ -- a unique value 
of total mass, $M_{\rm Ch}$ (the analogue of the Chandraseckhar limit). We find
\begin{equation}
M_{\rm Ch} = 6.67\times 10^{18}\left(\frac{m_f}{{\rm eV}}\right)^{-2}\,{\rm M}_{\odot}~.
\end{equation}

\subsection{On the existence of a lower bound for the Dark Matter mass}\label{sec:MassBound}

In ref.~\cite{Tremaine:1979we}, the possibility to extract lower bounds on the DM mass from the analysis 
of DM phase-space distribution in dwarf spheroidal galaxies 
has been discussed, referred to as the Tremaine-Gunn bound (see also refs.~\cite{Madsen:1983vg,Madsen:1991mz,Madsen:1990pe}). 
In ref.~\cite{Boyarsky:2008ju}, the special case of a degenerate fermionic 
self-gravitating gas has been explicitly discussed. 
To be more concrete, the strategy adopted in ref.~\cite{Boyarsky:2008ju} is the following.
For a spherically symmetric DM-dominated object with the mass $M$ within the region $R$, 
it is possible to obtain a lower bound on the DM mass by requiring that the Fermi velocity $v_{\rm F}$
of the degenerate gas of mass $M$ in the volume $V=4\pi R^3/3$ 
does not exceed the escape velocity $v_{\infty} = (2GM/R)^{1/2}$. More formally, this condition amounts to imposing the following inequality
\begin{equation}\label{eq:VelocityBound}
v_{\rm F} \leqslant v_{\infty} ~~\Rightarrow~~
\left[
\frac{3h^3\left(
\frac{3M}{4\pi R^3}
\right)}{8\pi m_f^4}
\right]^{1/3} 
\leqslant
\sqrt{
\frac{2GM}{R}
}~,
\end{equation}
where we used the mean density $\langle \rho\rangle=3M/4\pi R^3$.
Inverting for $m_f$, one obtains the bound \cite{Boyarsky:2008ju}
\begin{equation}\label{eq:NaiveBound}
m_f^4 \geqslant \frac{9h^3}{
64\sqrt{2}\pi^2 M^{1/2}G^{3/2}R^{3/2}
}~.
\end{equation}
Finally, using specific values for $R$ and $M$ extracted from observations,\footnote{In particular, in ref.~\cite{Boyarsky:2008ju} the value 
used for the radius is the half-light radius $r_{\rm half}$, i.e.\ the radius where the surface brightness profile falls to $1/2$ of its maximum value. 
The corresponding value of $M(r_{\rm half})$, on the contrary,
is estimated using the one-dimensional velocity dispersion.} it is possible -- for each one of the dwarf galaxy analyzed --
to convert eq.~(\ref{eq:NaiveBound})  into a numerical bound on $m_f$.
To give a taste, in ref.~\cite{Boyarsky:2008ju} the lower bound obtained, e.g., using the Carina dwarf spheroidal galaxy is $m_f > 215$ eV.\footnote{Stronger bounds can be derived assuming an initial thermal distribution of the DM particles~\cite{Boyarsky:2008ju}. As we will see in sec.~\ref{sec:Discussion}, this is not the case we will be interested in here and these bounds hence do not apply.}
The validity of this method clearly depends on the accuracy 
of the mass profile estimate.
In the case of a degenerate non-relativistic Fermi gas, moreover, we argue that the inversion of eq.~(\ref{eq:VelocityBound}) into eq.~(\ref{eq:NaiveBound})
cannot be performed straightforwardly, since both $R$ and $M$, as shown in eq.~(\ref{eq:R})
and eq.~(\ref{eq:M}),\footnote{These equation are written considering the whole degenerate configuration, i.e.\ for $\xi = \xi_1$. However, they still hold using a generic value of $\xi$, with $r = \xi\alpha$ and $M(r)=4\pi\int_0^r s^2\rho(s)ds$.} depend on the value of $m_f$ according to the scaling $R\sim 1/m_f^{4/3}$ and $M\sim 1/m_f^4$.
To be more concrete, considering
for definiteness 
the whole configuration, we find
\begin{eqnarray}
v_{\rm F}&\simeq& 8.2 \times 10^{12}
\left[
\frac{\rho_0}{{\rm kg/cm^3}}\left(\frac{{\rm eV}}{m_f}\right)^4
\right]^{1/3}\,{\rm km/s}~,\\
v_\infty&\simeq& 1.3 \times 10^{13}
\left[
\frac{\rho_0}{{\rm kg/cm^3}}\left(\frac{{\rm eV}}{m_f}\right)^4
\right]^{1/3}\,{\rm km/s}~,
\end{eqnarray} 
 with $v_{\rm F} < v_\infty$. As a consequence, in eq.~(\ref{eq:VelocityBound}) both the $m_f$- and $\rho_0$-dependence disappear, making eq.~(\ref{eq:NaiveBound}) useless if applied to the relevant parameter space of the model, i.e.\ $(m_f,\rho_0)$.
Nevertheless we can use the relation $v_\text{F} < v_\infty$ as a consistency check for our ansatz of a degenerate configuration. Instead of using eq.~(\ref{eq:VelocityBound}), we extract a bound on the parameter space imposing the condition
 \begin{equation}\label{eq:con}
 v_{\rm F} \leqslant v_{\infty}^{\rm obs}~,
 \end{equation}
 where the Fermi velocity, as discussed above, depends on the parameters $m_f$, $\rho_0$
 while $v_{\infty}^{\rm obs}$ is related to the velocity dispersion directly measured in astrophysical observations 
 via $v_{\infty}^{\rm obs} \simeq \sqrt{6}\sigma$,\footnote{
 The quantity directly observed is the 
 projection of the velocity of the stars along the line-of-sight, and it is a function of the projected radius (the so-called velocity dispersion, see eq.~(\ref{eq:LOSvelocity}) below).
 For DM-dominated objects -- like the dwarf spheroidal galaxies under scrutiny in this paper --
 rotation curves flatten, and the projected velocity can be characterized 
 by a single, constant value $\sigma$. We take the corresponding values from ref.~\cite{Walker:2009zp}. The relation $v_{\infty}^{\rm obs}\simeq \sqrt{6}\sigma$ holds under the assumption of isotropic velocity distributions.
 This approximation seems to be reasonable for DM particles \cite{Boyarsky:2008ju}, for which numerical simulations usually
predict a value of velocity anisotropy close to zero. As far as luminous stars are concerned, on the contrary, large values of 
velocity anisotropy are expected. We will discuss this issue in the next section.} where $\sigma$ is the one-dimensional velocity dispersion. We postpone to the next section 
 the exact definition of these quantities.
 
\section{Velocity dispersion of the Milky Way's classical dwarf spheroidal galaxies}\label{sec:velocity}

\begin{figure*}[!htb!]
\begin{center}
\centering
  \begin{minipage}{0.45\textwidth}
   \centering
   \includegraphics[width = \textwidth]{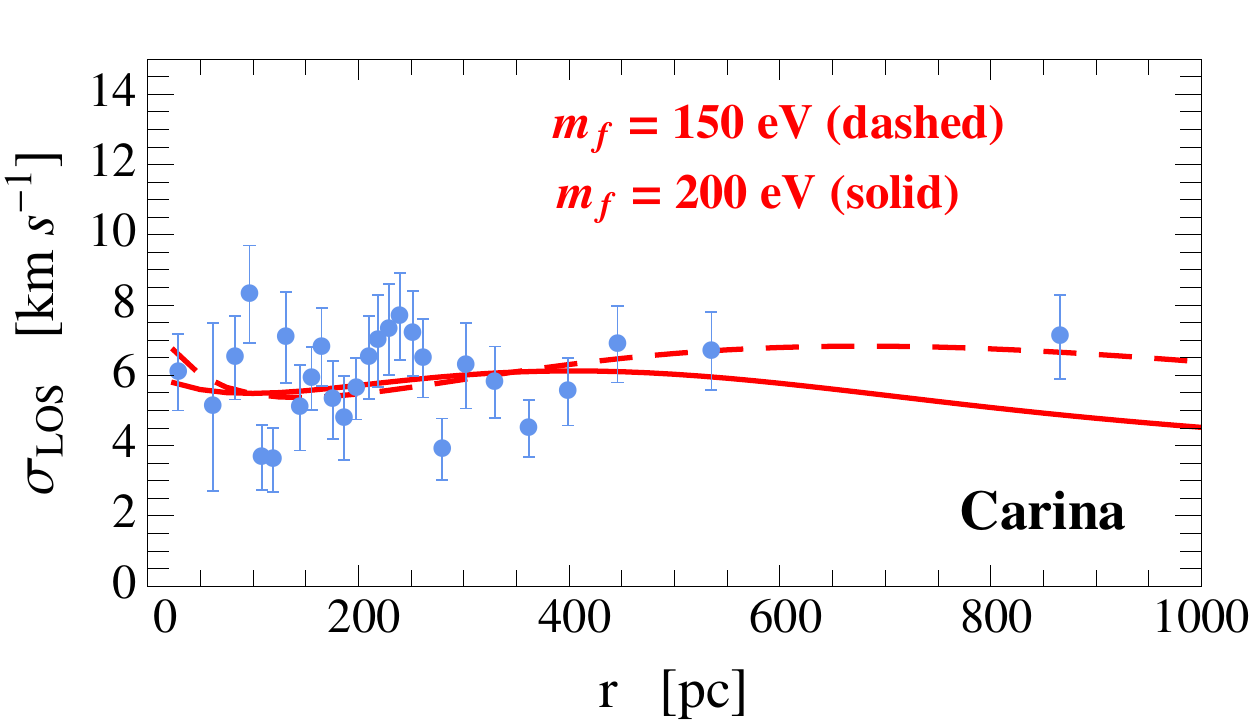}
    \end{minipage}\hspace{0.3 cm}
   \begin{minipage}{0.45\textwidth}
    \centering
    \includegraphics[width = \textwidth]{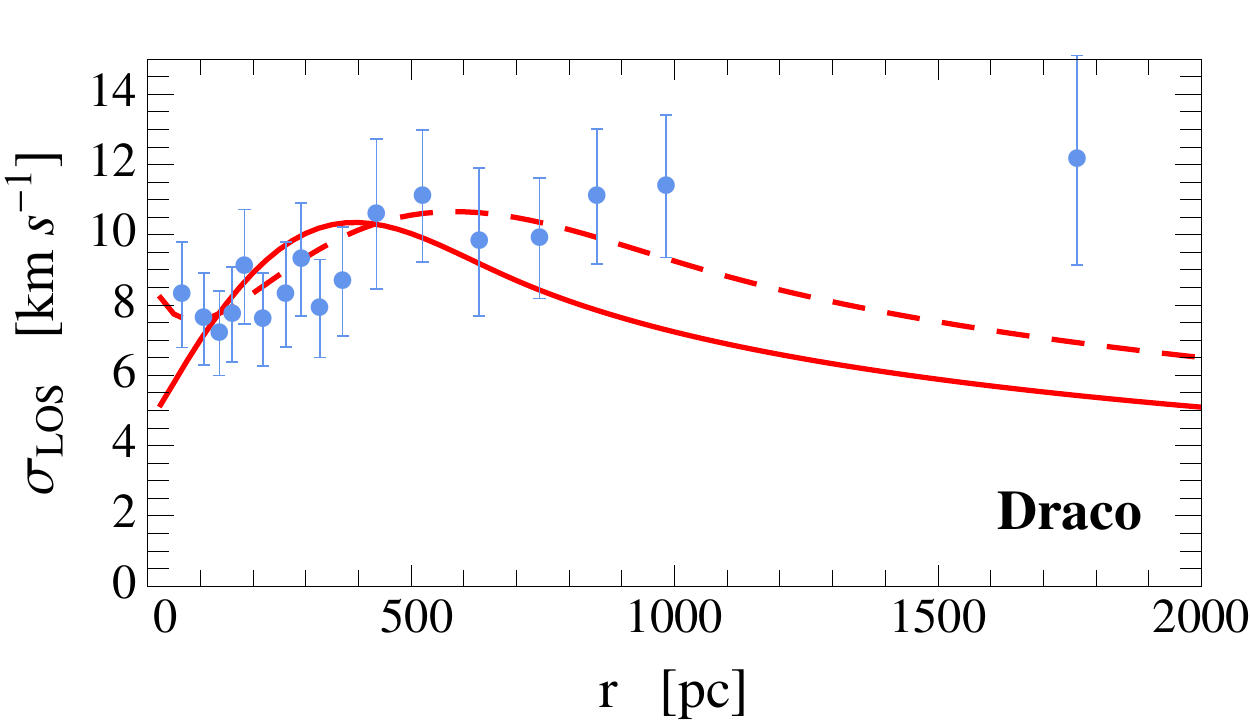}
    \end{minipage}\\
  \begin{minipage}{0.45\textwidth}
   \centering
   \includegraphics[width = \textwidth]{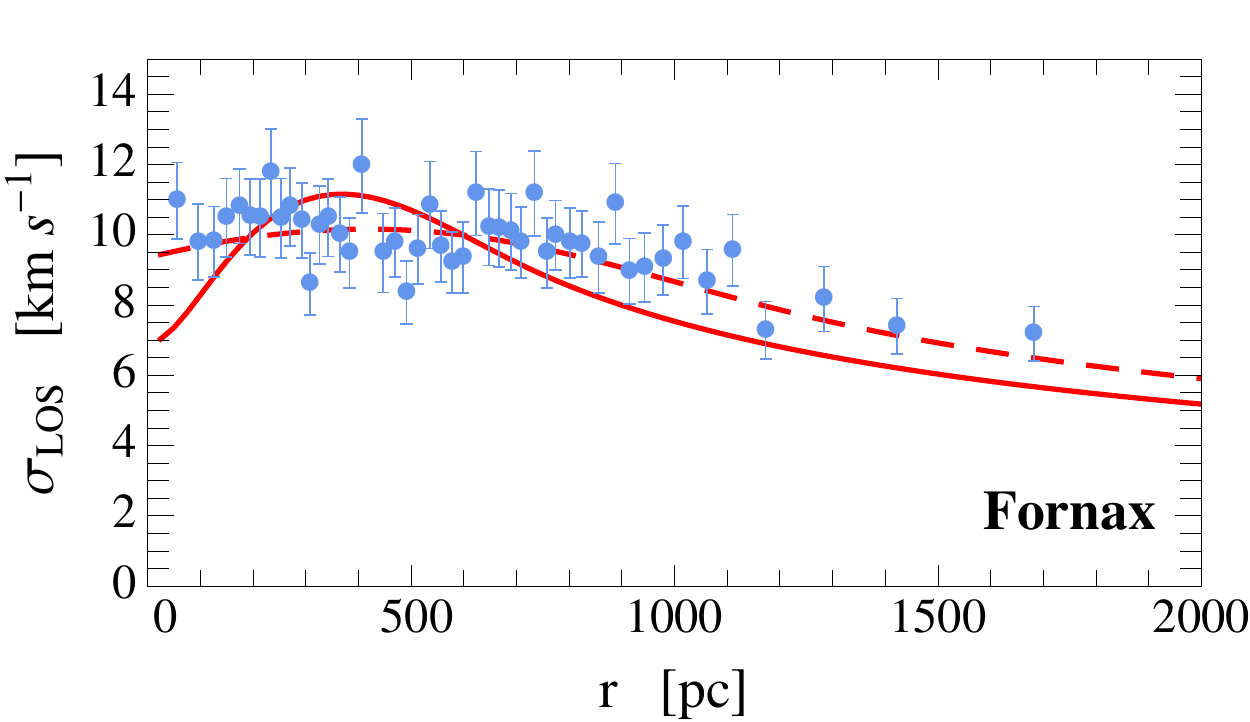}
    \end{minipage}\hspace{0.3 cm}
   \begin{minipage}{0.45\textwidth}
    \centering
    \includegraphics[width = \textwidth]{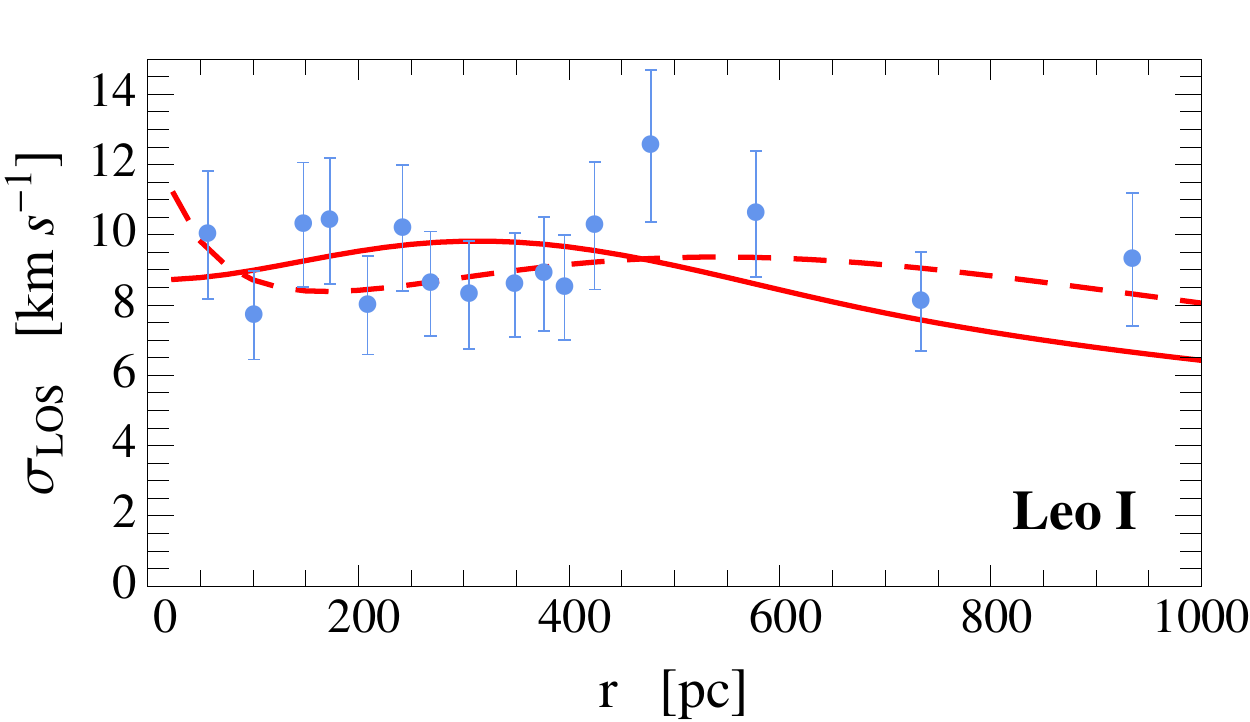}
    \end{minipage}\\
      \begin{minipage}{0.45\textwidth}
   \centering
   \includegraphics[width = \textwidth]{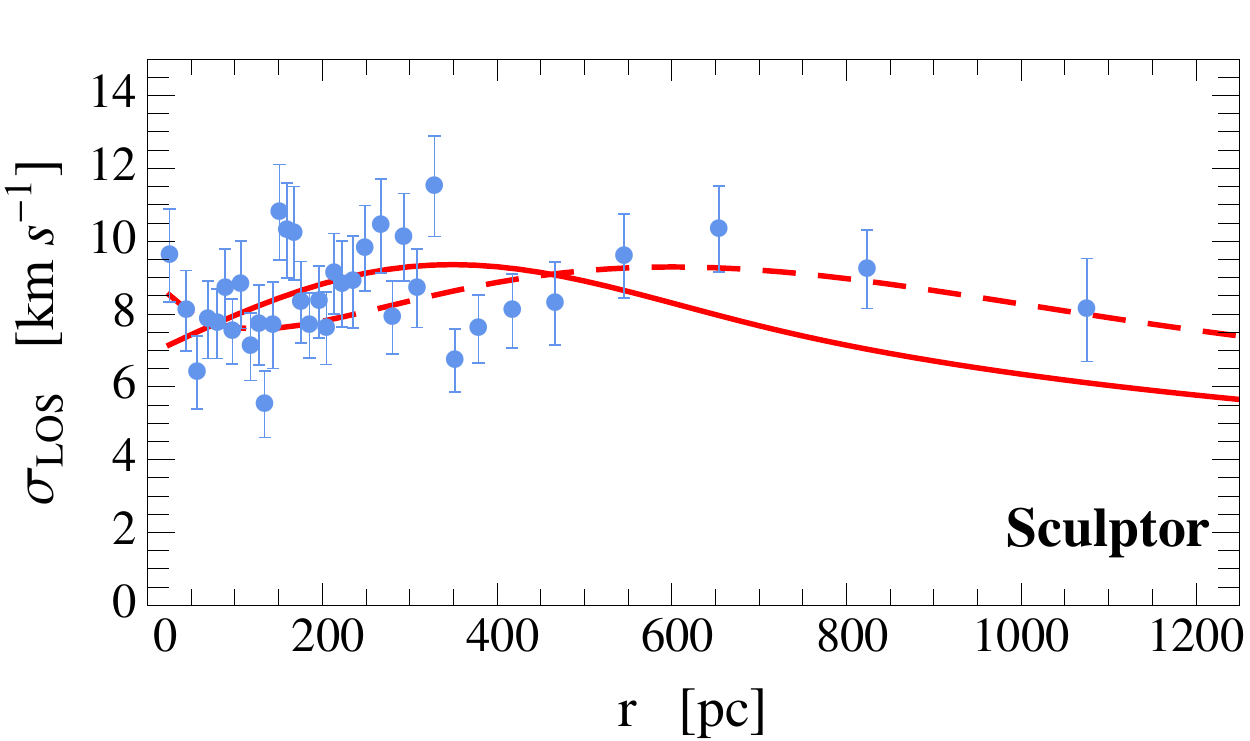}
    \end{minipage}\hspace{0.3 cm}
   \begin{minipage}{0.45\textwidth}
    \centering
    \includegraphics[width = \textwidth]{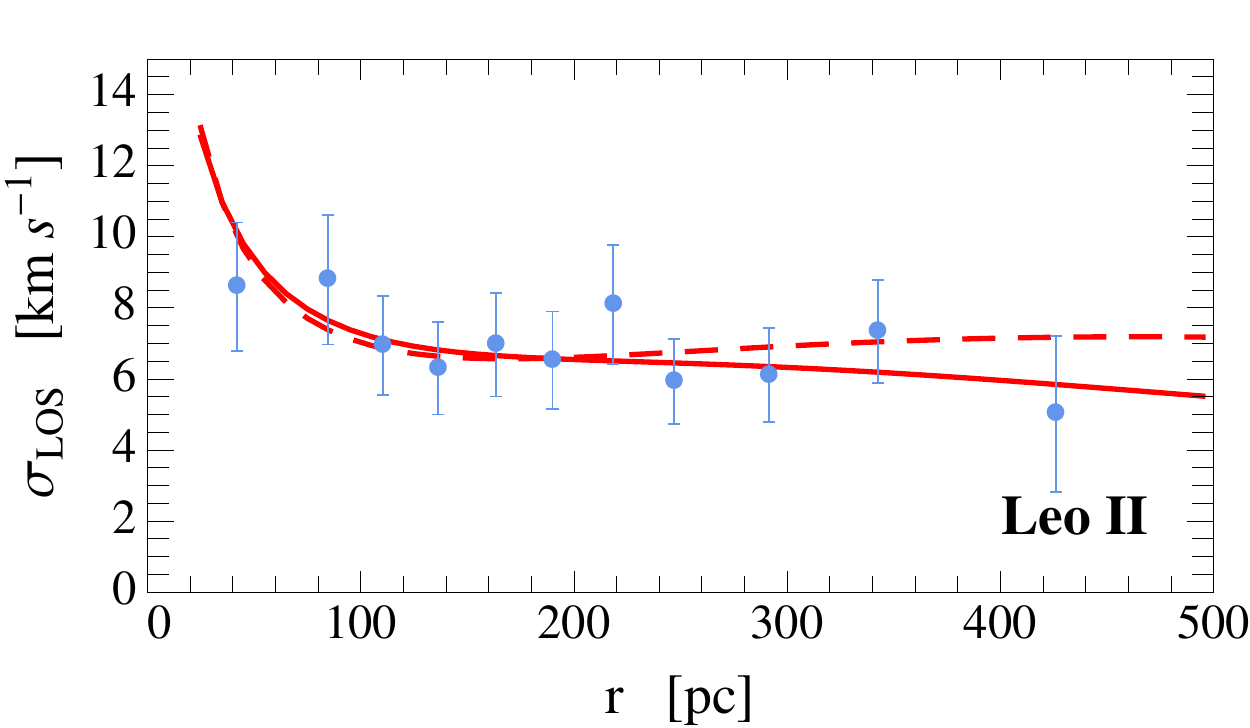}
    \end{minipage}
          \begin{minipage}{0.45\textwidth}
   \centering
   \includegraphics[width = \textwidth]{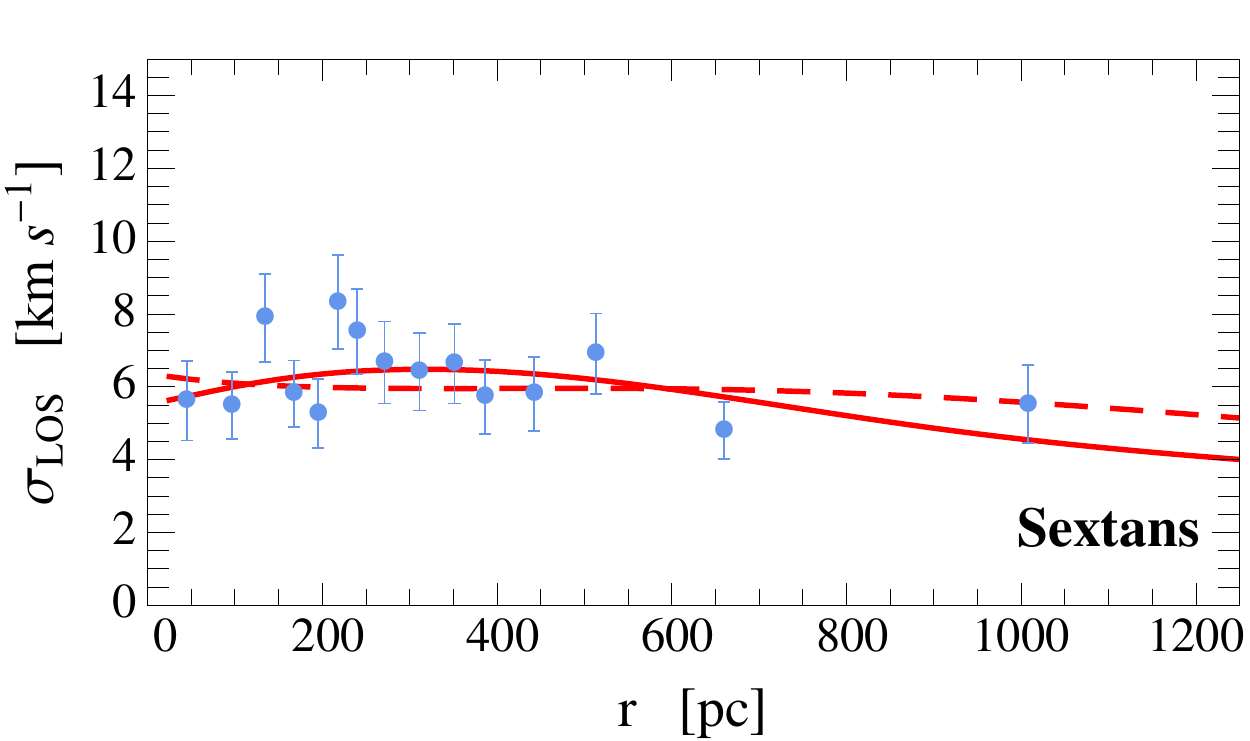}
    \end{minipage}\hspace{0.3 cm}
   \begin{minipage}{0.45\textwidth}
    \centering
    \includegraphics[width = \textwidth]{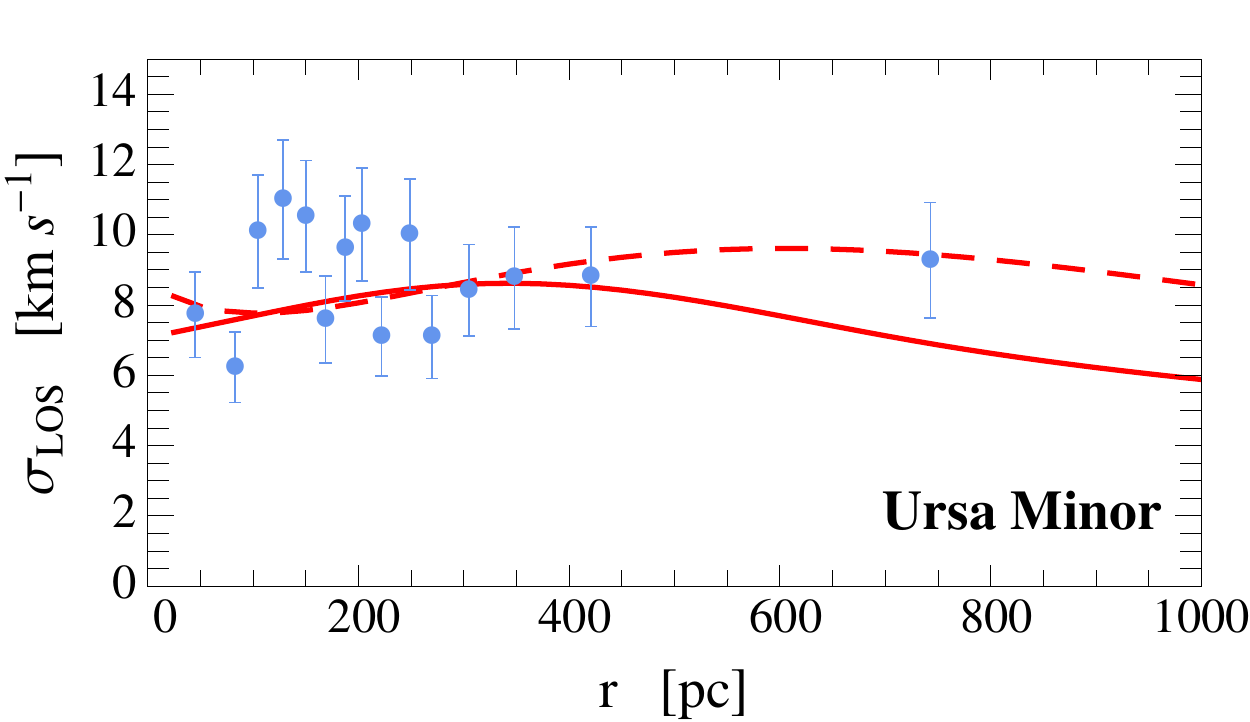}
    \end{minipage}
\caption{\textit{
Best fit of the 
projected velocity dispersion 
for the eight classical dwarf spheroidal galaxies. Data are taken from ref.~\cite{Walker:2009zp}. Here we show $m_f = 150$ eV (dashed line) and $m_f = 200$ eV (solid line). The details of the fit are given, respectively, in table~\ref{tab:Phi} and table~\ref{tab:Phi2}.
}}
\label{fig:ExSpectrum}
\end{center}
\end{figure*}

The Milky Way's dwarf spheroidal galaxies are low luminosity, low surface-brightness satellite galaxies characterized by no net rotation, 
a very large dynamical mass-to-light ratio and a small baryonic component. Eight of them, dubbed \textit{classical}, 
are characterized by 
high-quality data sets describing their stellar kinematic. 
In this section, we aim to use these data in order to test the degenerate Fermi gas model introduced in section~\ref{sec:FermiGasQ}.

In the analysis of the classical dwarf spheroidal 
galaxies only two quantities 
are directly observed: \textit{i}) the line-of-sight velocity dispersion 
as a function of the projected radius, and \textit{ii})
the surface brightness profile as a function of the  projected radius.
We refer the interested reader to ref.~\cite{Walker:2009zp} for a proper definition of these quantities. In the following we summarize, for the sake of clarity, the relevant 
formulas used in the fit.

The square of the projected velocity dispersion along the line-of-sight is
\begin{equation}\label{eq:LOSvelocity}
\sigma_{\rm LOS}^2(R) = \frac{2G}{I(R)}
\int_{R}^{\infty}
\nu(r^{\prime})
M(r^{\prime})(r^{\prime})^{2\beta - 2}
F(\beta, R, r^{\prime})dr^{\prime}~,
\end{equation}
where
\begin{equation}
F(\beta, R, r^{\prime}) \equiv
\int_{R}^{r^{\prime}}
\left(
1-\beta \frac{R^2}{r^2}
\right)\frac{r^{-2\beta + 1}}{
\sqrt{r^2 - R^2}
}dr~.
\end{equation}

We adopt the Plummer profile for the projected stellar density
\begin{equation}\label{eq:PlummerProfile}
I(R) = \frac{L}{\pi r_{\rm half}^2}\frac{1}{
[1+(R/r_{\rm half})^2]^2
}~,
\end{equation}
where $L$ is the total luminosity and $r_{\rm half}$ the half-light radius. 
We take the corresponding values from ref.~\cite{Walker:2009zp}.
The 3-dimensional stellar density, 
assuming spherical symmetry, is given by
\begin{eqnarray}
\nu(r) &=& 
-\frac{1}{\pi}\int_r^{\infty}
\frac{dI}{dr}
\frac{dR}{\sqrt{R^2 - r^2}}\nonumber \\
&=&
\frac{3L}{4\pi r_{\rm half}^3}
\frac{1}{
[1+(r/r_{\rm half})^2]^{5/2}}~.
\end{eqnarray}
Using in eq.~(\ref{eq:LOSvelocity})
the mass distribution obtained 
in section~\ref{sec:FermiGas}, we perform a $\chi^2$ fit of the
degenerate Fermi gas model against the velocity dispersion 
of the eight classical dwarf spheroidal galaxies of the Milky Way. We take the corresponding values from ref.~\cite{Walker:2009zp}.
For a fixed value of mass $m_f$, the model has only two free parameters, namely the central density $\rho_0$ and the orbital anisotropy of the stellar component $\beta$. The interesting feature of the model is that,  once $\rho_0$ and $\beta$ are fixed by the
fit to their best-fit values, all the properties of the DM halo are obtained consequently.

In fig.~\ref{fig:ExSpectrum}, we show our result for two benchmark values $m_f= 150$ eV and $m_f = 200$ eV. The details of the fit are collected in table~\ref{tab:Phi} and in table~\ref{tab:Phi2}. In both cases we obtain a decent fit, with $\chi^2_{\rm min}/{\rm d.o.f.}\simeq 1$ for all the classical dwarfs.
For the radius of the halo we obtain $R\simeq 1$ kpc, with $R\lesssim 1$ kpc ($R\gtrsim 1$ kpc) for $m_f = 200$ eV ($m_f = 150$ eV). 
As far as the total mass of the halo is concerned, we obtain 
$M \simeq 10^8$ M$_{\odot}$ (with, again, a preference for slightly smaller values if $m_f = 200$ eV). These values are in agreement 
with the estimate proposed in section~\ref{sec:FermiGas}, eq.~\eqref{eq:MRrelation}. 
Equipped with these results, we can compute for each dwarf galaxy the value of $M(r_{\rm half})$, i.e.\ the mass enclosed in a sphere of radius $r_{\rm half}$, 
predicted by the degenerate Fermi gas models. These values have been extracted in ref.~\cite{Walker:2009zp}
by means of a numerical Markov-chain Monte Carlo analysis. We compare the result of ref.~\cite{Walker:2009zp} with our predictions in table~\ref{tab:MassData}. 
Even if this comparison is not completely meaningful -- in ref.~\cite{Walker:2009zp} $M(r_{\rm half})$ is extracted assuming 
a specific halo profile (the generalized Hernquist profile, see also ref.~\cite{Hernquist:1990be}) -- it is nevertheless interesting to notice 
that we find a fairly good agreement. Let us now briefly comment about the values of the stellar anisotropy $\beta$ that we obtain from the fit.
On a general ground, it is possible to establish the following correspondences (see, e.g., ref.~\cite{Diez-Tejedor:2014naa}). 
 If all the stellar orbits are circular, $\beta = \infty$; if they are isotropic, $\beta = 0$; if they are perfectly radial, $\beta=1$; finally, tangentially biased systems correspond to $\beta < 0$. In principle, there is no a priori preference for anyone of these values; however, values $\beta \gtrsim 1$ seem to be disfavored by the 
 peculiar condition that they would require \cite{Diez-Tejedor:2014naa}. In our analysis we do not find any particular 
 preference for $\beta$; it ranges from $\beta = 1$ for Leo II if $m_f =150$ eV, to $\beta = -1.3$ for Fornax if $m_f = 200$ eV.

In fig.~\ref{fig:VelocityBound}, we compare the results of our analysis 
against the bound in eq.~(\ref{eq:con}). 
We plot the best-fit value for the central density $\rho_0$
as a function of the mass $m_f$, 
obtained marginalizing over the orbital anisotropy $\beta$.
The value $m_f = 150$ eV is consistent with the bound $v_{\rm F}\leqslant v_{\infty}^{\rm obs}$ for all the analyzed dwarf spheroidal galaxy but Leo II.
Note that this result has been obtained using a fixed profile for the stellar density, i.e.\ the Plummer profile in eq.~(\ref{eq:PlummerProfile}). This profile
is completely fixed once the value of $r_{\rm half}$ is specified. Different choices can be made (e.g.\ the King profile in ref.~\cite{KingProfile}), thus introducing extra 
free parameters that may change the result of the fit. 
The value $m_f = 200$ eV, on the contrary, is compatible with the bound in eq.~(\ref{eq:con}).
Since in this paper we are not interested in exploring more complicated setups for the stellar density, we identify the value $m_f \simeq 200$ eV as the best 
outcome of our phenomenological analysis. On a qualitative level, moreover, 
an upper bound on $m_f$ 
follows from the fact that larger values of $m_f$ imply smaller size for the DM halos, thus leading to unrealistic results.
To be more concrete, we find that even a value as large as $m_f \simeq 250$ eV leads to a significant worsening of the fit.
{Finally, in fig.~\ref{fig:HalfDensity} we show -- for all the eight classical dwarf spheroidal galaxies -- the values of the mean density within the half-light radius computed 
in our model considering $m_f \simeq 200$ eV and the corresponding best-fit values for $\rho_0$ shown in table~\ref{tab:Phi2}. 
We compare 
our models with the numerical results obtained in ref.~\cite{Walker:2009zp}; we note that the degenerate Fermi gas 
model seems to nicely reproduce the numerical data.  In particular, the  mean density within the half-light radius decreases 
for increasing values of $r_{\rm half}$.
} 

In summary, we find that the degenerate Fermi gas model can reproduce in a realistic way
 the kinematic data describing the velocity dispersion of the eight classical dwarf spheroidal galaxies of the Milky Way if $m_f \simeq 200$~eV.
Let us stress once again that this is a remarkable result given the simplicity of the model:
it provides a universal halo profile able to adapt -- for a nearly unique value of $m_f$ and with the only freedom of the structural parameter $\rho_0$ -- to all the observed kinematic data.
Furthermore,  
it has been obtained under the simplifying assumption that
 the whole system is in a perfect degenerate limit. The reader should keep in mind, in fact, that  in a more realistic situation 
 this is probably true only for a fraction of the DM particles.
 
{Let us close this section with a final remark.
One may wonder if the same conclusions obtained in this section can be reached without invoking quantum effects
 but instead considering the classical non-degenerate configuration of the gas. 
Interestingly, it is possible to show -- using the formalism developed in appendix~B -- that a description of the kinematic properties of the classical dwarf spheroidal galaxies
based on a classical isothermal gas leads to 
unrealistically large estimate for the size of their DM halo. 
The peculiar compact structure of these astrophysical objects, therefore, seems to point towards a configuration closed to the degenerate limit explored in this paper. }

 \section{Discussion: from the early Universe to the present day}\label{sec:Discussion} 

\subsection{Structure formation and Dark Matter abundance}\label{sec:FSL}

DM particles in the mass range identified in the last section, $m_f \simeq 200$~eV, are most likely produced as ultra-relativistic particles in the early Universe. Due to absence of interactions (apart from gravity) with the Standard Model (SM), they decouple immediately and are hence classified as  WDM.  As such their free-streaming length $\lambda_\text{FS}$ -- measuring the length-scale below which structures are erased -- must be checked against observations of the Lyman-$\alpha$ forest, of high- and low-z galaxies as well as high-z gamma ray bursts (see e.g.\ ref~\cite{Dayal:2014nva} and references therein for a recent analysis). To give a rough estimate, a free-streaming length above $\lambda \gtrsim 0.5$~Mpc is excluded, $0.3~\text{Mpc} \lesssim \lambda_\text{FS} \lesssim 0.5$~Mpc is an interesting window in which one might account for the deficit of structure on small scales compared to N-body simulations, and finally for $\lambda_\text{FS} < 0.1$~Mpc the particles behave as cold DM from the point of view of structure formation~\cite{Bringmann:2007ft}. In the simplest situation of a thermal relic, there is a one-to-one correspondence between $\lambda_\text{FS}$ and the DM mass, leading to the exclusion of WDM with a mass below $\sim 1 \div 2$ keV.

\begin{figure}
\centering
 \includegraphics[width = 0.45 \textwidth]{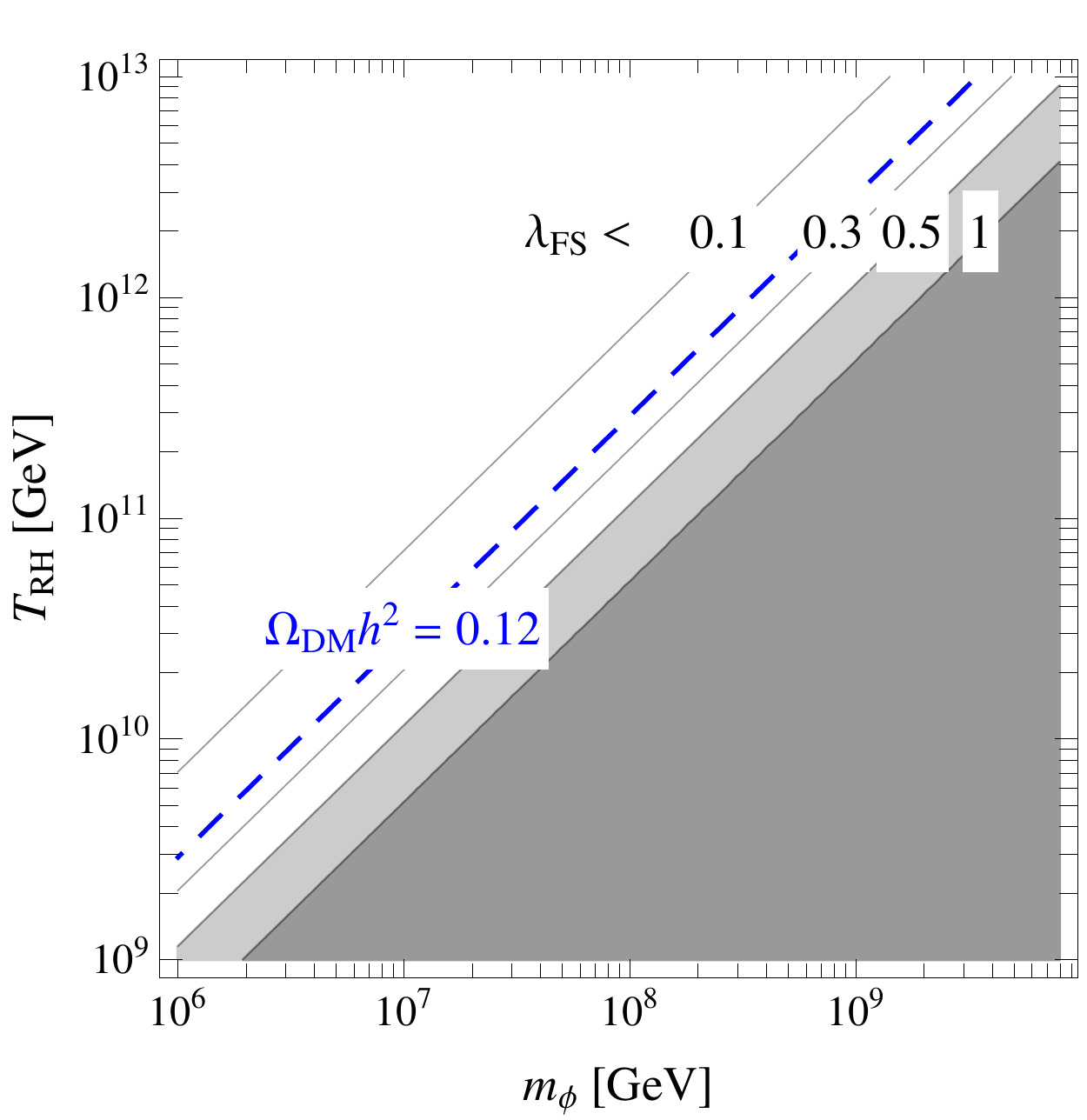}
\caption{ \textit{Bounds on the parameter space from the free-streaming length (gray shaded regions excluded) and the abundance of DM today (correct abundance obtained on dashed blue line) in terms of the inflaton mass $m_\phi$ and the reheating temperature $T_\text{RH}$ for $Br = 10^{-6}$ and $m_f = 200$~eV.}}
\label{fig_temp_mass}
\end{figure}

To retain our window of $m_f \sim 0.1 \div 0.3$~keV, we will in the following assume that the DM particles are produced non-thermally in the decay of the inflaton particle.\footnote{We focus on this production mechanism for simplicity. Note that other processes such as fermionic preheating~\cite{Greene:2000ew} can lead to a similar, equally predictive picture.} In this case, their initial momentum distribution is peaked a specific momentum $p_i = m_\phi/2$, where $m_{\phi}$ is the inflaton mass. Subsequently, the distribution is red-shifted as $p(t)/p_i = a_i/a(t)$, retaining the approximately monochromatic distribution. The free-streaming length can be calculated as
\begin{equation}
\begin{split}
 \lambda _{\rm FS} &= \int_{t_p}^{t_0} \frac{v}{a} dt \\
 &\simeq \frac{1}{ H_0 (1 + z_{eq})^{1/2} x} \ln \left[ x + \sqrt{1 +x^2} \right]  \,,
 \end{split}
\end{equation}
where $v(t) = p(t)/E(t)$ is the velocity of the DM particle, $x \equiv 4 \sqrt{t_{eq}} m_f/(\sqrt{t_p} m_\phi)$, $H_0 = 67.1$~km/(s Mpc) is the Hubble constant today, $z_{eq} = 3402$ is the redshift at matter-radiation equality and $t_p$, $t_{eq} = 0.695$~Myr and $t_0 = 13.8$~Gyr denote the time of DM production, of matter-radiation equality and today, 
respectively. Assuming a small branching ratio of the inflaton into DM particles, $Br = \Gamma_{\phi \rightarrow {\rm DM}}/\Gamma_\phi \ll 1$, the production time $t_p$ is determined by the reheating temperature $T_\text{RH}$
\begin{equation}
 t_p = \frac{1}{Br} \left( \frac{45}{\pi^2 g_{\gamma, RH}} \right)^{1/2} \frac{M_P}{T_\text{RH}^2}\,.
\label{eq:tp}
\end{equation}
Here $M_P = 2.4 \times 10^{18}$~GeV is the reduced Planck mass and $g_{\gamma}$ denotes the relativistic degrees of freedom contributing to the entropy of the photon bath, for the SM: $g_{\gamma,0} = 2$, $g_{\gamma, RH} = 427/4$, where the indices `RH' and `$0$' stand for the time of reheating and today, respectively. To obtain eq.~\eqref{eq:tp} we have exploited that the reheating temperature is determined by the decay rate into SM particles $\Gamma_{\phi \rightarrow \text{SM}} \simeq \Gamma_\phi$, $T_\text{RH} = (45/(\pi^2 g_{\gamma, {\rm RH}}))^{1/4} \sqrt{\Gamma_{\phi \rightarrow \text{SM} } M_P} $. In summary, this yields the free-streaming length as a function of the inflaton mass, the reheating temperature and the branching ratio, cf.\ fig.~\ref{fig_temp_mass}.

From the reheating temperature and the branching ratio we can calculate the DM abundance
\begin{equation}
 Y_\text{DM} = \frac{n_\text{DM}(T_\text{RH})}{s(T_\text{RH})} \simeq \frac{3}{4} Br \frac{T_\text{RH}}{m_\phi} \,,
\end{equation}
where we have used $\rho_\phi = m_\phi n_\phi \simeq \pi^2/30 \, g_{\gamma, {\rm RH}} T^4_\text{RH}$. The observed abundance today~\cite{Ade:2013zuv} implies
\begin{equation}
 0.12 \stackrel{!}{=} \Omega_\text{DM} h^2 = 41.1 \, Br \frac{T_\text{RH}}{m_\phi} \frac{m_f}{200 \text{ eV}} \,.
\end{equation}
Together with the constraint from the free-streaming length, $\lambda_\text{FS} < 0.5$, this implies
\begin{equation}
 Br < 4.9 \times 10^{-6} \,, \quad m_\phi/T_\text{RH} < 1.3 \times 10^{-3} \,,
\label{eq:Tmbounds}
\end{equation}
 for our reference value of $m_f = 200$~eV. 
The viable parameter space of our model covers both small values of $\lambda_\text{FS}$ in which structure forms as in the cold DM paradigm, as well as larger values (close to the bounds in eq.~\eqref{eq:Tmbounds}), in which the observed deficit of structure on small scales might be explained. Moreover, it is indeed remarkable that the hypothesis of a degenerate fermion gas as main component of dwarf galaxies has lead us to conclusions on the preceding inflation model, even more so as these constraints can indeed be fulfilled in some typical inflation models.

 \subsection{A criterium for degeneracy}\label{sec:tdeg}

In the analysis of section~\ref{sec:velocity} we have assumed that dwarf galaxies have reached a degenerate configuration. Given the production of the DM particles in the early Universe at high energies, this implies the occurrence of a phase transition. A comprehensive study of this problem
 is of course beyond the purposes of this paper; nevertheless it is interesting to provide at least some simple arguments in favor of this possibility.\footnote{See ref.~\cite{Chavanis:2002rj} for a general overview of the problem.}
 For a gas of fermions, the degeneracy temperature $T_{\rm DEG}$ is defined as follows\footnote{The inclusion of the gravitational potential into the distribution function does not alter this expression; see eq.~\eqref{eq:FDStatistic} in appendix~\ref{app:FermiDirac} and the subsequent discussion.}
\begin{eqnarray}\label{eq:TDEG}
T_{\rm DEG} &=& \frac{h^2}{2\pi m_f k_{\rm B}}\left(\frac{\rho_0}{2m_f}\right)^{2/3} \\
&=& 1.7 \times 10^{-3} \, K \left( \frac{\rho_0}{10^{-27} \frac{\text{kg}}{\text{cm}^3} } \right)^{2/3} \left( \frac{200 \text{ eV}}{m_f} \right)^{5/3} \!, \nonumber
\end{eqnarray}
 where $k_{\rm B}$ is the Boltzmann constant. To good approximation, this corresponds to the Fermi temperature of a gas of non-relativistic particles: $T_{\rm F} = E_{\rm F}/k_{\rm B}$, $E_{\rm F} = p_{\rm F}^2/(2 m_f)$.
 If the present temperature of the fermionic gas is much lower than this degeneracy temperature, 
 then a necessary condition to have a degenerate configuration is satisfied. 

To address the question of consistency of such a degenerate configuration today, let us have a brief look at its cosmological history. Let us consider a  particle of mass $m_f \simeq 200$~eV with no sizable couplings to the  SM, produced relativistically in the early Universe in the 2-body decay of e.g.\ the inflaton particle. As described above, the resulting momentum distribution is strongly peaked. This changes only when structure formation sets in at a redshift of $z = {\cal{O}}(1 - 10)$.\footnote{The effect of this on the calculation of the free-streaming length above is negligible, since there the dominant contribution arises from $t < t_{eq}$.} Due to their gravitational interaction, local overabundances of DM density (seeded by primordial fluctuations) form virialized DM halos. The time-dependent gravitational forces lead to violent relaxation, converting the non-thermal peaked distribution into an equilibrium Fermi-Dirac distribution within a few dynamical times~\cite{Padmanabhan1, LyndenBell:1966bi,Chavanis:2002rj,Chavanis:2002yv,Chavanis:2002D}. The corresponding temperature can be estimated using the virial theorem equating the total kinetic and potential energies,
\begin{equation}
\begin{split}
 T_{\text{DM},0} &\sim \frac{G M m_f}{R \ k_\text{B}}\,\\
&=  10^{-2} \, K \, \frac{M}{10^8\,{\rm M}_{\odot}} \frac{\text{kpc}}{R} \frac{m_f}{200~\text{eV}}\,.
\end{split}
\end{equation}
For typical values for the mass and radius of dwarf galaxies, cf.\ fig.~\ref{fig:MassRadioRelation}, this estimate yields a value close to the degeneracy temperature. On the other hand, for large galaxies we find $T_\text{DEG} \simeq 10^{-4}\,  K \ll T_{\text{DM},0} \sim 10^{-1} \,K$, where we have used $m_f = 200$~eV, $M = 10^{11} M_{\odot}$, $R = 50$~kpc, $\rho \sim  M/(\frac{4}{3} \pi R^3)$ as a reference. This estimate  shows that temperatures below the degeneracy limit are indeed within reach for dwarf galaxies, while larger galaxies might be described by a thermal distribution of the same DM particle, which is however not in the degenerate limit. This simple argument thus supports the picture sketched in refs.~\cite{Destri:2013pt,deVega:2013woa,deVega:2013jfy}. In the next section, we will discuss this point in more details.

Finally, there is one more consistency check we can perform on the cosmological history of our DM particle. Ignoring structure formation, we can calculate the `would-be' temperature $\widetilde T_{\text{DM}}$ of the DM particles today, based solely on the redshift of their peaked distribution. Since the gravitational collapse during structure formation heats up the DM gas, a necessary condition is $\widetilde T_{\text{DM}} < T_{\text{DM},0}$. From $p \propto a^{-1}$ we find
\begin{equation}
 \begin{split}
  &E_k \sim k_{\rm B} T \sim p/c\phantom{m^2} ~~\Rightarrow~~T \propto a^{-1} \text{   for } T \gg m_f\,, \\
  &E_k \sim k_{\rm B} T \sim p^2/2m~~\Rightarrow~~T \propto a^{-2} \text{   for } T \ll m_f\,,
 \end{split}
\label{eq:Ta}
\end{equation}
where $E_k$ is the kinetic energy.
Based on eq.~\eqref{eq:Ta} we can estimate the would-be temperature of the DM sector today as
\begin{equation}
 \widetilde T_{\rm DM} \simeq \frac{k_{\rm B}\,  m_\phi^2}{4\,  m_f} \left( \frac{T_{\gamma,0}}{T_{\text{RH}}} \right)^2 \left( \frac{g_{\gamma, 0}}{g_{\gamma, i}}\right)^{2/3}\,,
 \label{eq:T0}
\end{equation}
where the subscripts $0$ and $i$ stand for today and an early initial time, respectively. $T_\gamma$ denotes the temperature of the photon bath with $T_{\gamma,0} = 2.7$~K. For $m_f = 200$~eV, the condition $\widetilde T_{\text{DM}} < T_{\text{DM},0}$ is fulfilled as long as $m_\phi \lesssim 10^2 \, T_\text{RH}$. From fig.~\ref{fig_temp_mass} we see that for the values of $m_\phi$ and $T_\text{RH}$ in accordance with bounds from structure formation, this is easily fulfilled: indeed $m_\phi \lesssim 10^{-4} \, T_\text{RH}$.

\subsection{Larger galaxies and the non-degenerate configuration}\label{sec:LargerGalaxies}

We have mentioned the possibility that, within the framework studied in this paper, larger galaxies correspond to non-degenerate configurations of the gas. This picture has been studied in the context of WDM in refs.~\cite{Destri:2013pt,deVega:2013woa,deVega:2013jfy}.
However, since we are considering a significantly different value of the DM mass, their conclusions 
do not apply straightforwardly to our case. 
The aim of this section is to provide a simple quantitative analysis able to attest the validity of the aforementioned hypothesis.  

The statistical analysis of a self-gravitating Fermi gas at non-zero temperature
can be carried out in analogy to what was discussed in 
section~\ref{sec:FermiGas} for the special case of the degenerate configuration.
We review the basic formalism in appendix~\ref{app:FermiDirac}.
In a nutshell, it is possible to retrace the same discussion
outlined in section~\ref{sec:FermiGas} using the Fermi-Dirac distribution at finite temperature
 instead of the  degenerate limit in eq.~\eqref{eq:FDT0}. It turns out that  in a non-degenerate configuration the gas is characterized by the following equation
\begin{equation}\label{eq:NonDegGas}
\frac{8\pi \sqrt{2}m_f^{5/2}(k_{\rm B}T)^{3/2}}{h^3}=\frac{\rho_0}{\mathcal{I}_{1/2}(k)}~.
\end{equation} 
This equation relates the temperature $T$, the central density $\rho_0$, and the dimensionless parameter $k$ which controls the 
degree of degeneracy of the gas.\footnote{The definition of the Fermi integral $\mathcal{I}_{n}(k)$ in eq.~(\ref{eq:NonDegGas}) can be found in appendix~\ref{app:GeneralFD}.} The limit $k\to \infty$ corresponds to 
the classical limit of an isothermal gas described by the Maxwell-Boltzmann statistic, while the limit $k\to 0$ corresponds to the Fermi degenerate gas (see appendix~\ref{app:GeneralFD}).
For a given value of $k$ the mass density $\rho(r)$ of the system can be obtained by numerically solving 
a generalized Lane-Emden equation (see eq.~(\ref{eq:PoissonRew})), and  the central value $\rho_0$ -- or, equivalently,  the temperature via eq.~(\ref{eq:NonDegGas}) -- is a free parameter 
that we need to extract from observations. 
Our approach is the following.
As customary in the analysis involving the Burkert DM profile \cite{Burkert:1995yz}
we define the core radius $R_{\rm H}$ as the radius 
where the DM density equals one fourth of its central value.
The corresponding core mass is $M_{\rm H}=\int_0^{R_{\rm H}}
4\pi r^2\rho(r) dr$.
We compute both these quantities 
for different values of $k$ and $\rho_0$, while we keep the DM mass fixed at $m_f = 200$ eV.
 \begin{figure}[b]
\centering
 \includegraphics[width = 0.45 \textwidth]{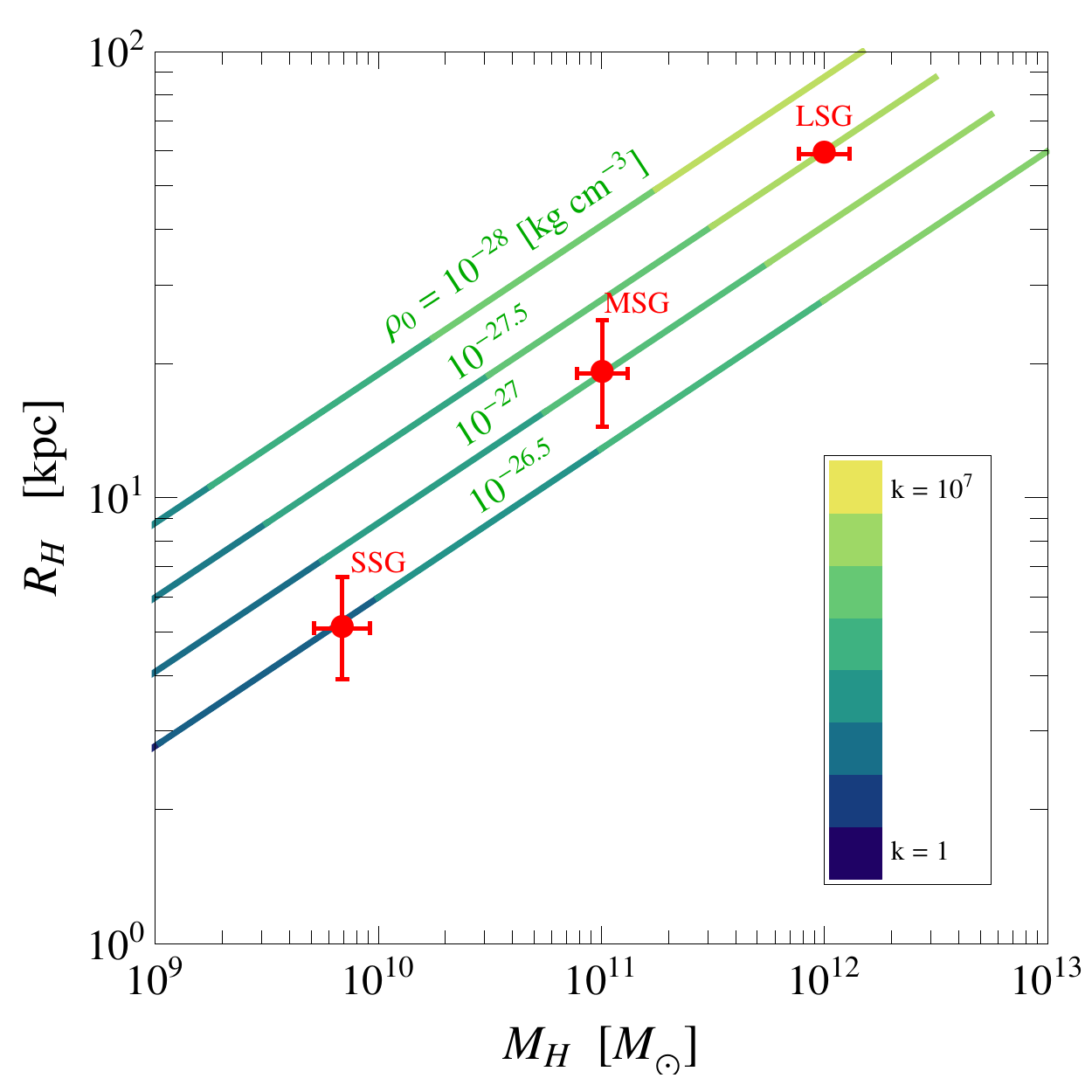}
\caption{ \textit{
Core mass-radius relation for the non-degenerate Fermi gas against data describing small, medium and large spiral galaxies. The DM mass is fixed to $m_f = 200$ eV.
We plot the theoretical prediction of the model for four different values of the central density $\rho_0$, while on each curve 
the gradient color marks different values of $k$. The limit $k\to \infty$ correspond to the Maxwell-Boltzmann regime.
}}
\label{fig:LargerGalaxies}
\end{figure}
In fig.~\ref{fig:LargerGalaxies} we compare our results 
against  
the characteristic values of core mass and radius for small spiral galaxies (SSG), medium spiral galaxies (MSG) 
and large spiral galaxies (LSG). These characteristic data are taken from ref.~\cite{Destri:2013pt}.
We plot the theoretical prediction 
for the core mass-radius relation 
of the non-degenerate Fermi gas considering four different values of central density, from $\rho_0 = 10^{-28}$ kg/cm$^3$ to
$\rho_0 = 10^{-26.5}$ kg/cm$^3$, while the color gradient spans the range $k = 1\div 10^7$.
The model can easily accomodate large galactic structures for increasing value of $k$, thus confirming the general 
picture in which 
large galaxies are described by the Maxwell-Boltzmann limit of the theory. To be more concrete, 
we find that SSG correspond to $k\sim \mathcal{O}(10)$ and $T \sim 0.1$ K,
MSG to $k\sim \mathcal{O}(10^3)$ and $T \sim 0.5$ K, and LSG to $k\sim \mathcal{O}(10^6)$ and $T\sim 1$ K.

Before concluding, let us discuss one more interesting point. In ref.~\cite{Donato:2009ab}
it has been noticed 
that the observed surface density $\Sigma_0 \equiv R_{\rm H}\rho_0$
is nearly constant for the observed galaxies and does not depend on the galaxy luminosity.
This is a remarkable result, 
in particular since 
it has been obtained analyzing 
different 
galactic systems in a range that covers over 14 magnitudes in luminosity. The best-fit value of the surface density 
obtained in ref.~\cite{Donato:2009ab} is $\Sigma_0 = 141^{+81}_{-52}$ M$_{\odot}$/pc$^2$.
We can compare the prediction of the non-degenerate Fermi model extracted from our fig.~\ref{fig:LargerGalaxies} against this value. We find $\Sigma_0 \simeq 238$ M$_{\odot}$/pc$^2$ (SSG),
$\Sigma_0 \simeq 280$ M$_{\odot}$/pc$^2$ (MSG), $\Sigma_0 \simeq 275$ M$_{\odot}$/pc$^2$ (LSG). 
Therefore, the value of surface density predicted by the model 
is nearly constant, even if slightly larger w.r.t. the one observed in ref.~\cite{Donato:2009ab} (but still compatible within the errors). The reader should keep in mind, moreover, that our analysis is not based on real data
but only on an order-of-magnitude estimate for the core mass and radius of spiral galaxies.
A more detailed analysis is mandatory but it is beyond the purpose of this paper.

In conclusion, we found that the picture according to which 
large galaxies correspond to the non-degenerate limit of the Fermi gas
is consistent in our model with $m_f= 200$ eV.\footnote{{
More general -- yet not less motivated -- scenarii are possible. 
For instance, an interesting possibility 
is to consider a galaxy made of 
an inner core of almost constant density governed by degenerate quantum statistics surrounded 
by a large external part fulfilling the classical Maxwell-Boltzmann statistics.
The possibility to test this scenario, as well as the general picture provided by the model analyzed in this paper, requires
a numerical study of the relaxation process during galaxy formation. This is beyond the scope of the present paper, and it will be 
explored in forthcoming studies.} 
}

\section{Conclusions}\label{sec:Conclusions}

{
Numerical simulations with cold DM deviate from
 observations at galactic scales. 
Two of the most problematic aspects are related to 
the prediction of cuspy (rather than cored) density profiles for dwarf galaxies, and to an
overwhelming abundance of small structures that are not observed.
Motivated by these problems, in this paper we proposed a simple
alternative paradigm in which DM is made of free fermions with mass $m_f$. 
We can summarize the most tantalizing consequences of 
this assumption as follows.

In the first part of the paper, we started our analysis from the Universe that we observe today. 
\begin{itemize}
\item Describing a galactic structure as a self-gravitating Fermi gas of DM particles, 
we focused our attention on the case of dwarf spheroidal galaxies.
In particular, we assumed that a dwarf spheroidal galaxy  corresponds to the degenerate limit of the gas, 
where the attractive force of gravity is entirely balanced by the quantum pressure arising from the Pauli exclusion principle.
In this picture, therefore, dwarf spheroidal galaxies are quantum astrophysical objects. 

\item 
In the degenerate configuration DM halos are described by a cored mass density profile.
We tested 
the model against the kinematic data describing the velocity dispersion of the eight classical dwarf spheroidal galaxies of the Milky Way. We 
found a good agreement with data providing that $m_f \simeq 200$~eV.  We pointed out why this value is not in violation of the Tremaine-Gunn bound.

\item Larger galaxies correspond to non-degenerate configurations of the Fermi gas. We tested this 
picture for the value $m_f \simeq 200$~eV. We showed that large spiral galaxies correspond to the classical Maxwell-Boltzmann limit, while small spiral galaxies are closer to the degenerate configuration.

\end{itemize}

Going back in time, we analyzed the implications of the value $m_f \simeq 200$ eV
throughout the history of the Universe. 

\begin{itemize}

\item We discussed a concrete realization in which DM is produced non-thermally by the decay of the inflaton
during reheating.
This mechanism  offers a remarkable connection between the primordial Universe and the Universe observed today.
DM particles are ultra-relativistic at the time of their production  
but they have to be non-relativistic at the time of matter-radiation equality, 
since otherwise the structures that we observe today would be erased.

\item 
By computing the free-streaming length,
imposing the bound 
from the Lyman-$\alpha$ forest, and 
requiring to reproduce the correct value of DM relic density,
 we obtained a consistent picture only for specific values of the inflaton mass, reheating temperature and branching ratio  for the decay of the inflaton into DM.
 Moreover, we also noticed that in a region of the allowed parameter space the model can easily explain
 the deficit of small structures observed today.

\end{itemize}

To sum up, 
starting from the assumption that dwarf spheroidal galaxies are quantum astrophysical objects
we presented a simple model of fermionic DM  consistent with observations. 
Most importantly, 
the model features a remarkable connection between the early and the present Universe that can be tested by constraining the parameters of the early Universe (e.g.\ by a future measurement of the stochastic gravitational wave background of inflation or of the amplitude of the primordial B-modes in the CMB) or by
 improving the current  kinematic description of Milky Way's dwarf spheroidal galaxies.
 In particular, it would be interesting to perform the same analysis outlined in this paper for the faintest 
 dwarf spheroidal galaxies
 for which velocity dispersion profiles are not yet available.

}

Open questions remain. In particular, it would be interesting to extend our qualitative arguments for larger galaxies by a more quantitative study, focusing on the value of $m_f$ found here. Furthermore, the impact of the constraints found here on inflation model building remain to be analyzed in more detail. We leave these questions to future work. 

\smallskip

\acknowledgments

We thank Marco Cirelli, Mauro Valli, Wei Xue and in particular Piero Ullio for discussions and important advice. 
A.U. is especially grateful to Felix Br\"ummer, Sacha Davidson, Miha Nemev\v{s}ek and Kenichi Saikawa for many inspiring discussions during the axion workshop at IPNL. Moreover, we thank Felix Br\"ummer for collaboration on the inital stages of this project and Sacha Davidson for valuable comments on the draft.
This work is supported by the ERC Advanced Grant n$^{\circ}$ $267985$, ``Electroweak Symmetry Breaking, Flavour and Dark Matter: One Solution for Three Mysteries" (DaMeSyFla) (A.U.) and by the European Union FP7-ITN INVISIBLES (Marie Curie Action PITAN-GA-2011-289442-INVISIBLES) (V.D.).

\appendix

\section{Fit of the velocity dispersion}\label{app:Fit}

In this appendix we collect the numerical results 
of the fit performed in section~\ref{sec:velocity}. Table~\ref{tab:Phi} refers to $m_f = 150$ eV, while table~\ref{tab:Phi2}
to $m_f = 200$ eV. In table~\ref{tab:MassData} we compare the values of $M(r_{\rm half})$ obtained in ref.~\cite{Walker:2009zp} with the prediction 
of the degenerate Fermi gas model.

\begin{table*}[!htb!]
\fbox{\footnotesize $m_f = 150$ eV}\vspace{0.2 cm}
\centering
\begin{tabular}{|c||c|c|c||c|c|c|c|c|}\hline
   dSphs & {\color{red}{$\rho_0$ [$10^{-27}$ kg cm$^{-3}$]}} & {\color{red}{$\beta$}} & $\chi^2_{\rm min}/{\rm d.o.f.}$ &
    {\color{blue}{$R$ [kpc]}} & {\color{blue}{$M$ [$10^8$ M$_{\odot}$]}}  & 
    {\color{blue}{$\langle \rho \rangle$ [$10^{-28}$ kg cm$^{-3}$]}} & 
   {\color{blue}{$v_{\rm F}$ [km s$^{-1}$]}} & $v_{\infty}$ [km s$^{-1}$]  \\ \hline\hline
    \textbf{Carina} & $3.1$ & $0.46$ & $1.41$ & $1.41$ & $0.9$ & $4.7$ & $14.4$ & $16.2$   \\ \hline
    \textbf{Draco} & $10$ & $0.3$ & $0.55$ & $1.16$ & $1.6$ & $8.4$ & $17.6$ &  $22.3$  \\\hline
    \textbf{Fornax} & $3.9$ & $-0.1$ & $0.65$ & $1.35$ & $1.0$ & $4.0$ & $15.1$ & $28.6$  \\ \hline
    \textbf{Leo I} & $7.9$ & $0.5$ & $0.57$ & $1.21$ & $1.4$ & $7.5$ & $17.0$ & $22.5$  \\ \hline
    \textbf{Sculptor} & $6.3$ & $0.3$ & $1.42$ & $1.25$ & $1.3$ & $6.7$ & $16.3$ & $22.5$   \\ \hline
    \textbf{Leo II} & $10$ & $1$ & $0.40$ & $1.16$ & $1.6$ & $8.4$ & $17.6$ & $16.1$ \\ \hline
    \textbf{Sextans} & $1.0$ & $0.1$ & $0.97$ & $1.70$ & $0.5$ & $2.6$ & $12.0$ & $19.3$  \\ \hline
    \textbf{Ursa Minor} & $6.3$ & $0.2$ & $1.34$ & $1.25$ & $1.3$ & $6.7$ & $16.3$ & $23.3$  \\\hline
                        \end{tabular}\vspace{0.3cm}
                        \caption{\label{tab:Phi} \textit{
   Result of the analysis of the degenerate Fermi gas model
   against data describing the velocity dispersion of the eight classical dwarf spheroidal galaxies, for $m_f = 150$ eV. 
  The two columns in red represent the best-fit values obtained for the central density $\rho_0$ and the anisotropy parameter $\beta$.
  The columns in blue represent, for the best-fit values of $\rho_0$ and $\beta$, the predictions of the model, namely the total mass $M$ and radius $R$
  of the DM halo, the mean mass density $\langle \rho\rangle$, and the Fermi velocity $v_{\rm F}$.
For comparison, in the last column we report an estimate for the escape velocity 
  obtained as $v_{\infty} \simeq \sqrt{6}\sigma$ (the values of $\sigma$ are taken from ref.~\cite{Walker:2009zp}).
 }}
     \end{table*}
     
\begin{table*}[!htb!]
\fbox{\footnotesize $m_f = 200$ eV}\vspace{0.2 cm}
\centering
\begin{tabular}{|c||c|c|c||c|c|c|c|c|}\hline
   dSphs & {\color{red}{$\rho_0$ [$10^{-27}$ kg cm$^{-3}$]}} & {\color{red}{$\beta$}} & $\chi^2_{\rm min}/{\rm d.o.f.}$ &
    {\color{blue}{$R$ [kpc]}} & {\color{blue}{$M$ [$10^8$ M$_{\odot}$]}}  & 
    {\color{blue}{$\langle \rho \rangle$ [$10^{-28}$ kg cm$^{-3}$]}} & 
   {\color{blue}{$v_{\rm F}$ [km s$^{-1}$]}} & $v_{\infty}$ [km s$^{-1}$]  \\ \hline\hline
    \textbf{Carina}         & $4.8$ & $0.18$ & $1.50$ & $0.89$ & $0.35$ & $1.8$ & $7.2$ & $16.2$   \\ \hline
    \textbf{Draco}          & $15.8$ & $-0.7$ & $1.48$ & $0.73$ & $0.65$ & $3.4$ & $8.8$ &  $22.3$  \\\hline
    \textbf{Fornax}        & $12.6$ & $-1.3$ & $2.03$ & $0.76$ & $0.57$ & $3.0$ & $8.5$ & $28.6$  \\ \hline
    \textbf{Leo I}           & $15.8$ & $0.0$ & $0.87$ & $0.73$ & $0.65$ & $3.4$ & $8.8$ & $22.5$  \\ \hline
    \textbf{Sculptor}      & $12.6$ & $-0.2$ & $1.73$ & $0.76$ & $0.57$ & $3.0$ & $8.5$ & $22.5$   \\ \hline
    \textbf{Leo II}          & $15.8$ & $0.9$ & $0.35$ & $0.73$ & $0.65$ & $3.3$ & $8.8$ & $16.1$ \\ \hline
    \textbf{Sextans}      & $2.5$ & $-0.3$ & $0.81$ & $0.99$ & $0.26$ & $1.3$ & $6.5$ & $19.3$  \\ \hline
    \textbf{Ursa Minor} & $10.0$ & $-0.1$ & $1.44$ & $0.79$ & $0.51$ & $2.7$ & $8.2$ & $23.3$  \\\hline
                        \end{tabular}\vspace{0.3cm}
                        \caption{\label{tab:Phi2} \textit{
The same as in table~\ref{tab:Phi}, but for $m_f = 200$ eV.
 }}
     \end{table*}     
     
\begin{table}[!htb!]
\centering
\begin{tabular}{|c||c|c|c|}\hline
    \multirow{3}{*}{dSphs} & $M(r_{\rm half})$ &  $M(r_{\rm half})$ & $M(r_{\rm half})$  \\ 
      &  [$10^7$ M$_{\odot}$]  &  [$10^7$ M$_{\odot}$]  &  [$10^7$ M$_{\odot}$]  \\
      & from ref.~\cite{Walker:2009zp}\footnote{In ref.~\cite{Walker:2009zp}, $M(r_{\rm half})$ is estimated in two ways: using a numerical Markov chain Monte Carlo method and by means of a simple analytic model based on the Jeans equation. Since the latter assumes, contrary to our analysis, a stellar velocity 
      distribution that is isotropic $\beta = 0$, we compare the prediction of the degenerate Fermi model with the value of $M(r_{\rm half})$
      extracted in ref.~\cite{Walker:2009zp} from the full numerical analysis.} & $m_{f} = 150$ eV & $m_{f} = 200$ eV  \\ \hline\hline
    \textbf{Carina} & $0.4^{+0.1}_{-0.1}$  & 0.25 & 0.36 \\ \hline
    \textbf{Draco} & $0.6^{+0.5}_{-0.3}$ & $0.44$ & $0.64$ \\\hline
    \textbf{Fornax} & $4.3^{+0.6}_{-0.7}$ & $4.6$ & $5.6$ \\ \hline
    \textbf{Leo I} & $1.0^{+0.6}_{-0.4}$ & $0.67$ & $1.2$ \\ \hline
    \textbf{Sculptor} & $1.0^{+0.3}_{-0.3}$ & $0.63$ & $1.1$  \\ \hline
    \textbf{Leo II} & $0.5^{+0.2}_{-0.3}$ & $0.20$ & $0.31$ \\ \hline
    \textbf{Sextans} & $1.6^{+0.4}_{-0.4}$ & $1.4$ & $2.0$  \\ \hline
    \textbf{Ursa Minor} & $1.3^{+0.3}_{-0.5}$ & $0.77$ & $1.1$  \\\hline
                        \end{tabular}\vspace{0.3cm}
                        \caption{\label{tab:Mass} \textit{
  Comparison between  the values of  $M(r_{\rm half})$ predicted by the degenerate Fermi model
  using the best fit analysis in table~\ref{tab:Phi} and table~\ref{tab:Phi2},
  and the values of $M(r_{\rm half})$ obtained in ref.~\cite{Walker:2009zp}.
 }}\label{tab:MassData}
     \end{table}     
     
\begin{figure*}[!htb!]
\begin{center}
\centering
  \begin{minipage}{0.45\textwidth}
   \centering
   \includegraphics[width = \textwidth]{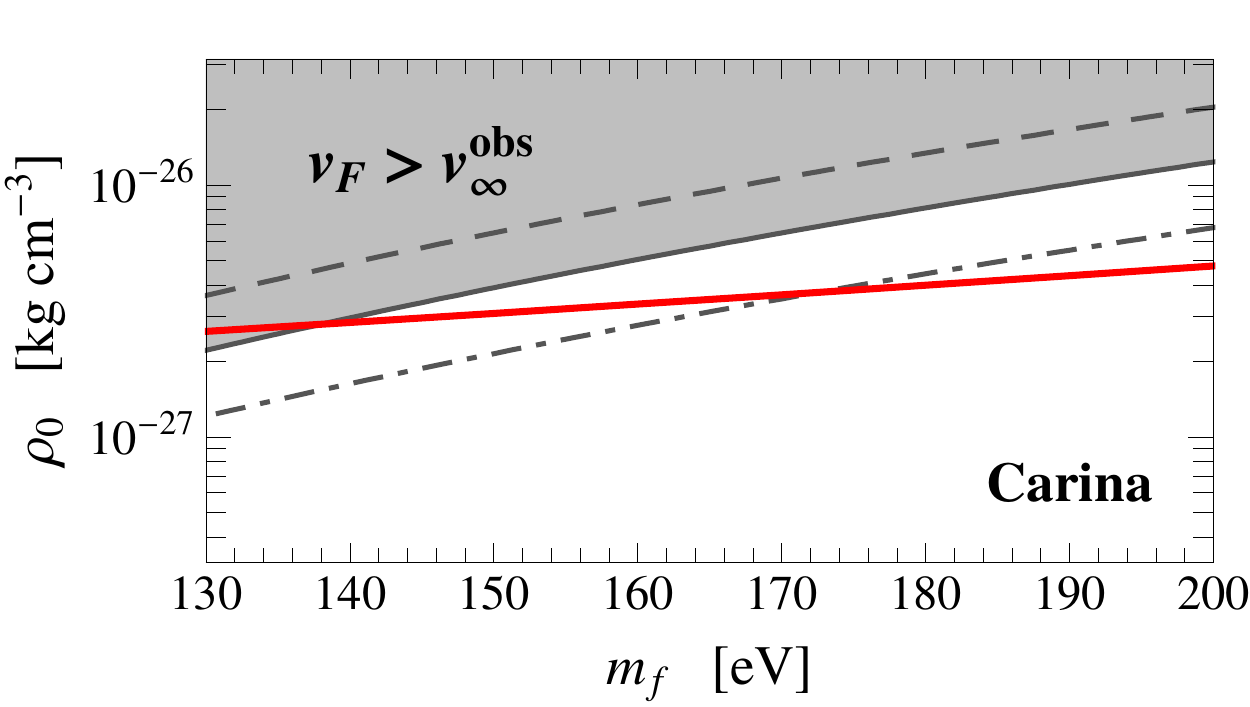}
    \end{minipage}\hspace{0.3 cm}
   \begin{minipage}{0.45\textwidth}
    \centering
    \includegraphics[width = \textwidth]{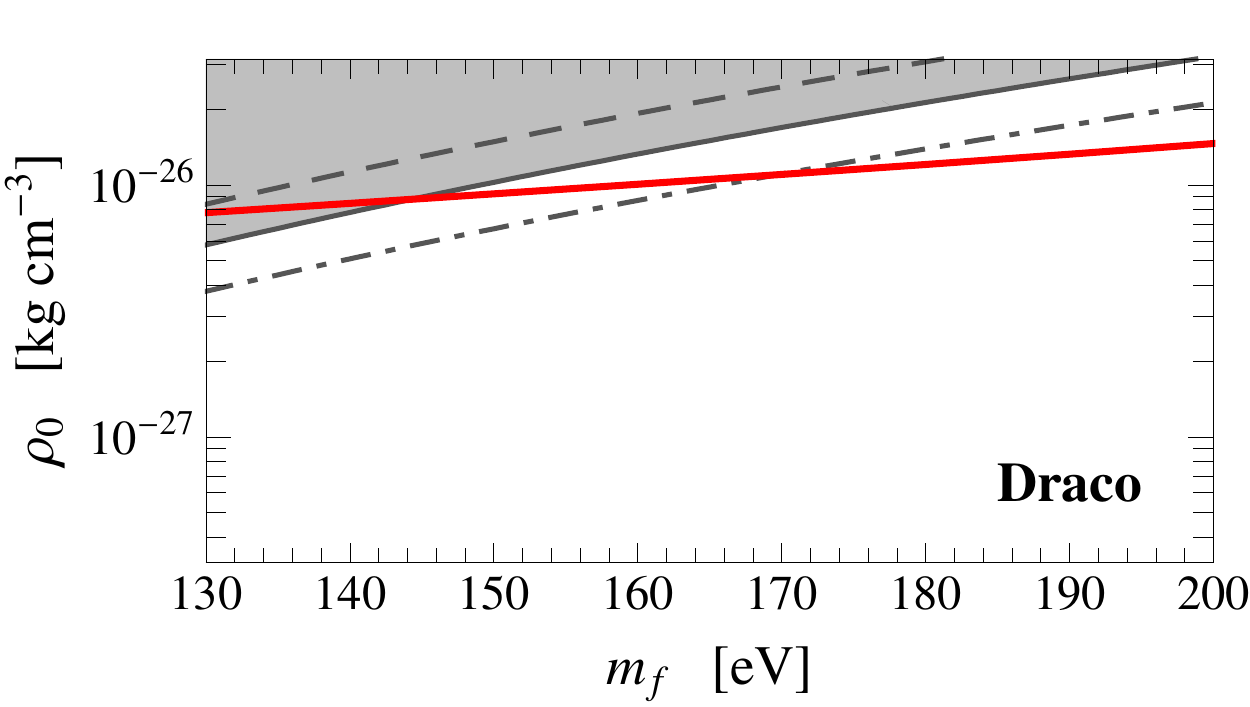}
    \end{minipage}\\
  \begin{minipage}{0.45\textwidth}
   \centering
   \includegraphics[width = \textwidth]{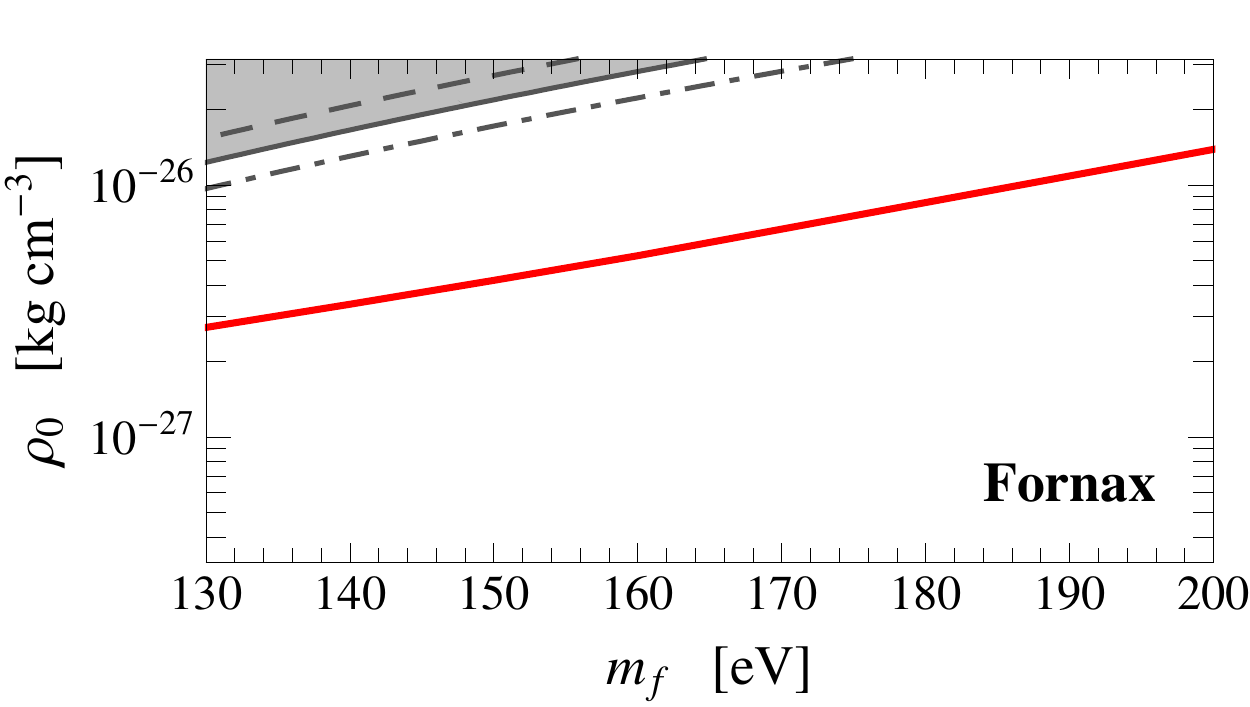}
    \end{minipage}\hspace{0.3 cm}
   \begin{minipage}{0.45\textwidth}
    \centering
    \includegraphics[width = \textwidth]{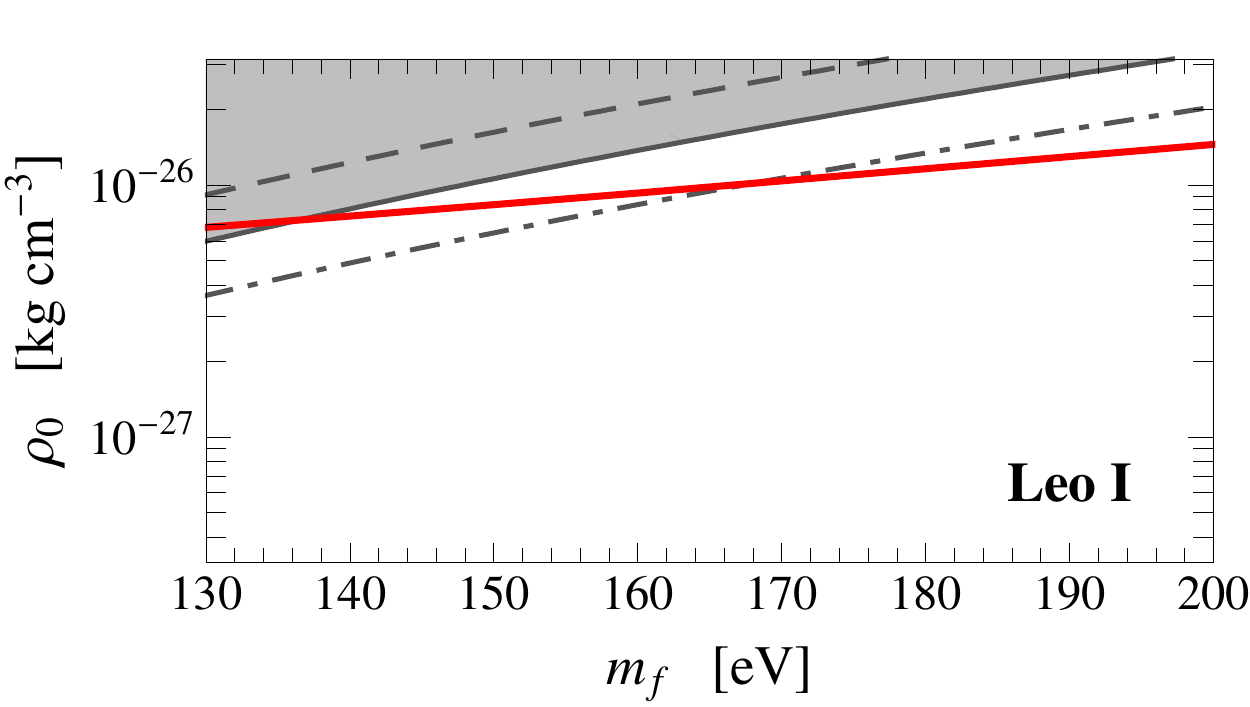}
    \end{minipage}\\
    \begin{minipage}{0.45\textwidth}
   \centering
   \includegraphics[width = \textwidth]{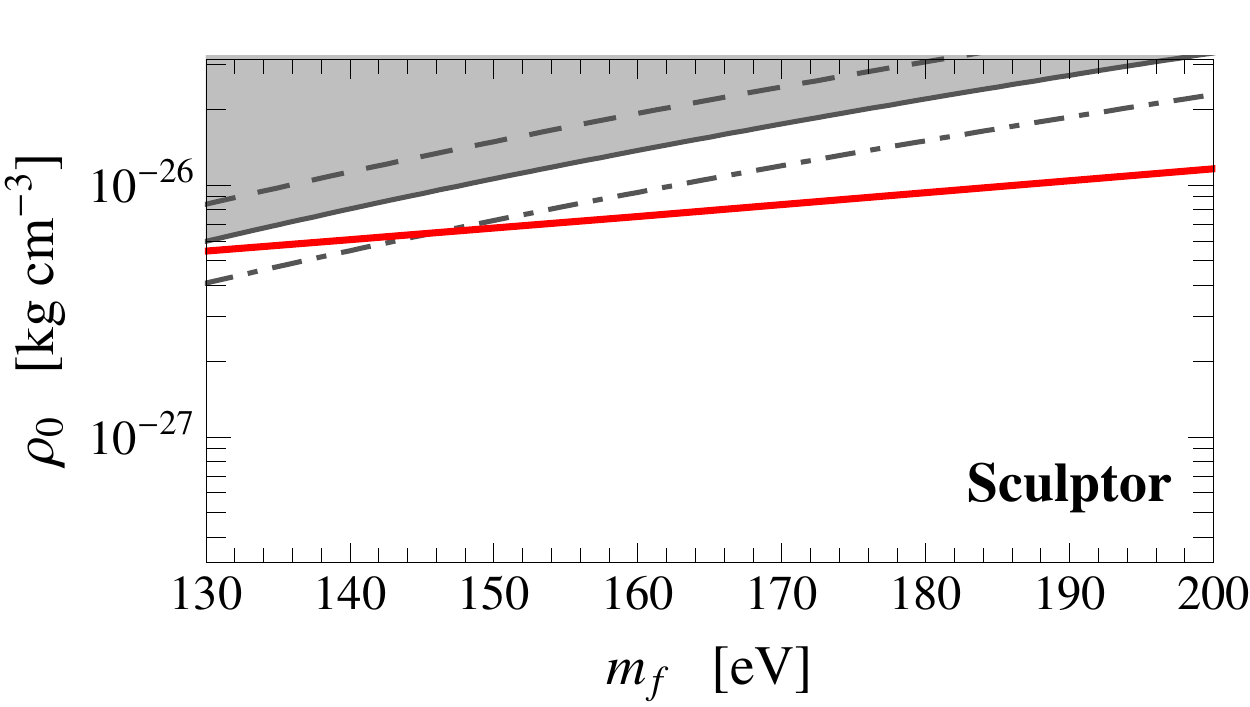}
    \end{minipage} \hspace{0.3 cm}
   \begin{minipage}{0.45\textwidth}
    \centering
    \includegraphics[width = \textwidth]{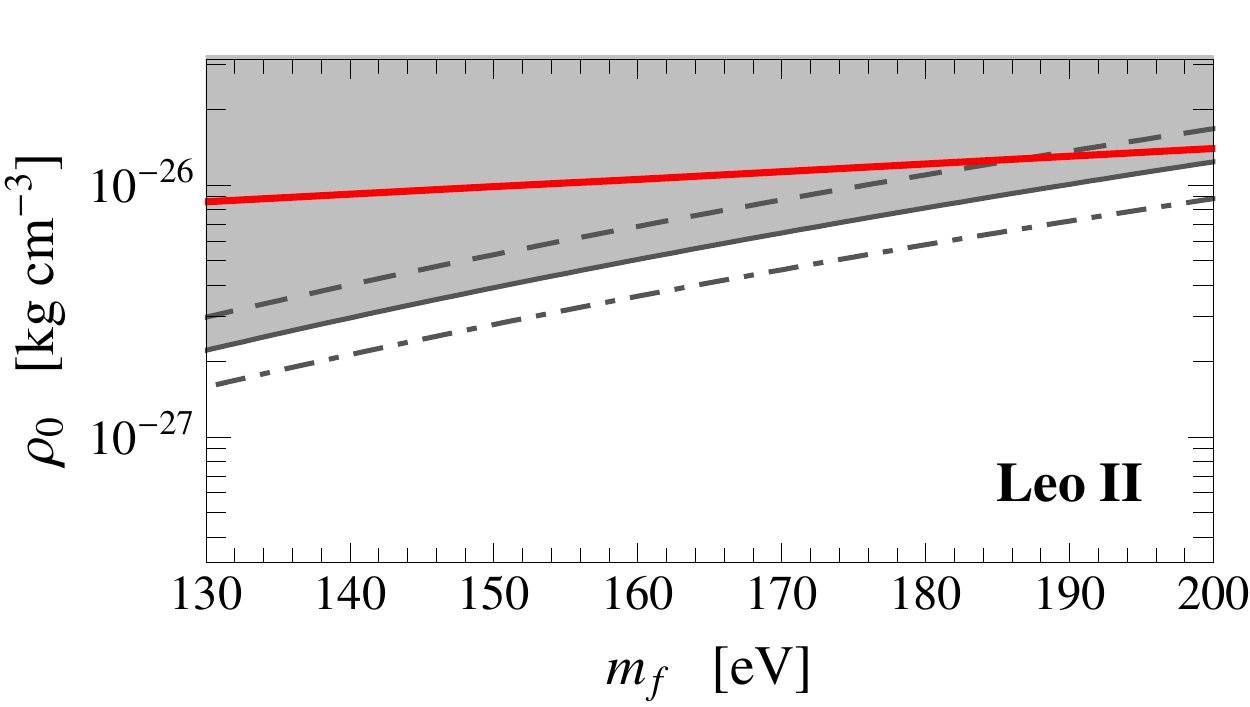}
    \end{minipage}\\
        \begin{minipage}{0.45\textwidth}
   \centering
   \includegraphics[width = \textwidth]{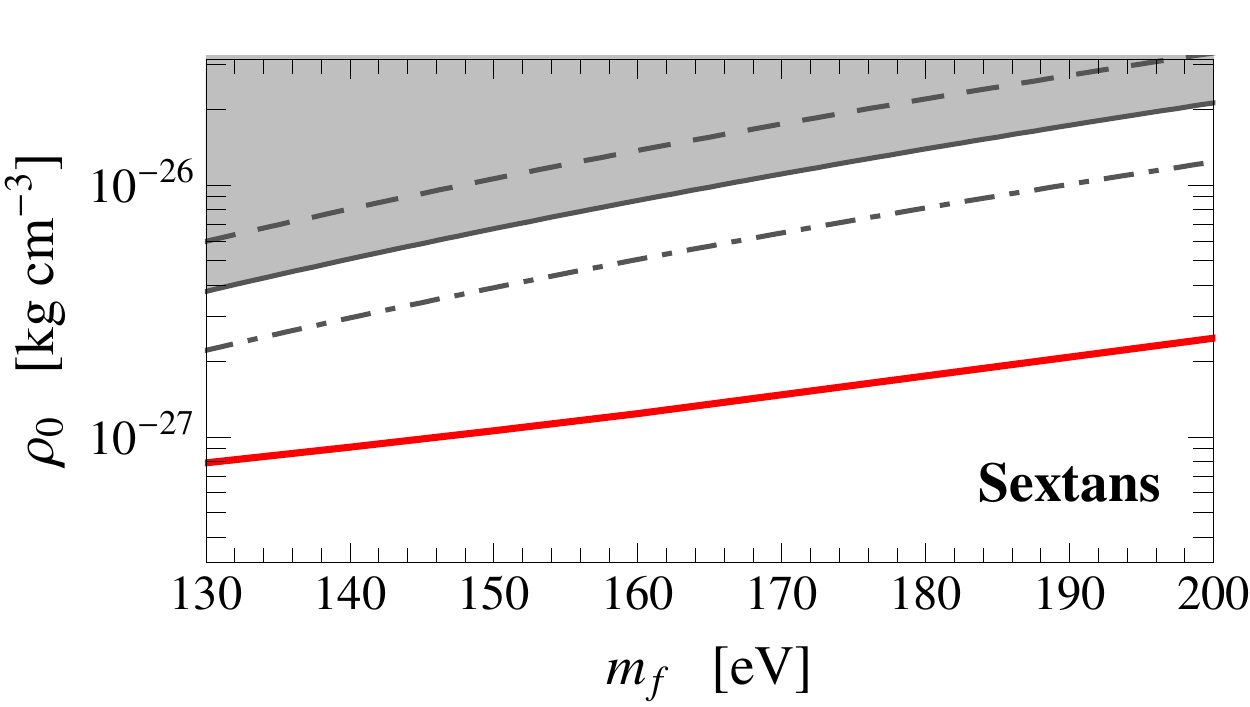}
    \end{minipage} \hspace{0.3 cm}
   \begin{minipage}{0.45\textwidth}
    \centering
    \includegraphics[width = \textwidth]{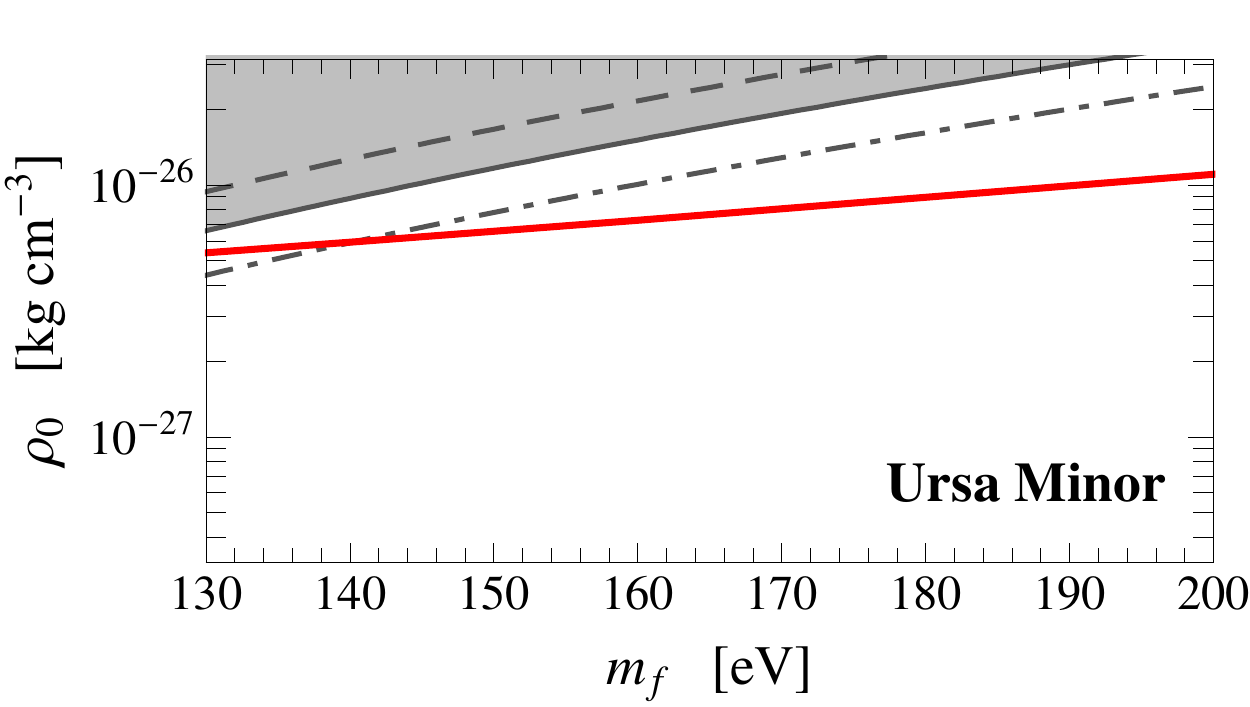}
    \end{minipage}
\caption{\textit{
Best-fit value for the central density $\rho_0$ (solid red)
 as a function of the mass $m_f$, obtained marginalizing over $\beta$. 
 The  gray shaded region corresponds to the condition imposed in eq.~(\ref{eq:con}).
 Dashed and dot-dashed lines account for the error in $v_{\infty}^{\rm obs}$. }}
\label{fig:VelocityBound}
\end{center}
\end{figure*}

 \begin{figure}[!htb!]
\centering
 \includegraphics[width = 0.47 \textwidth]{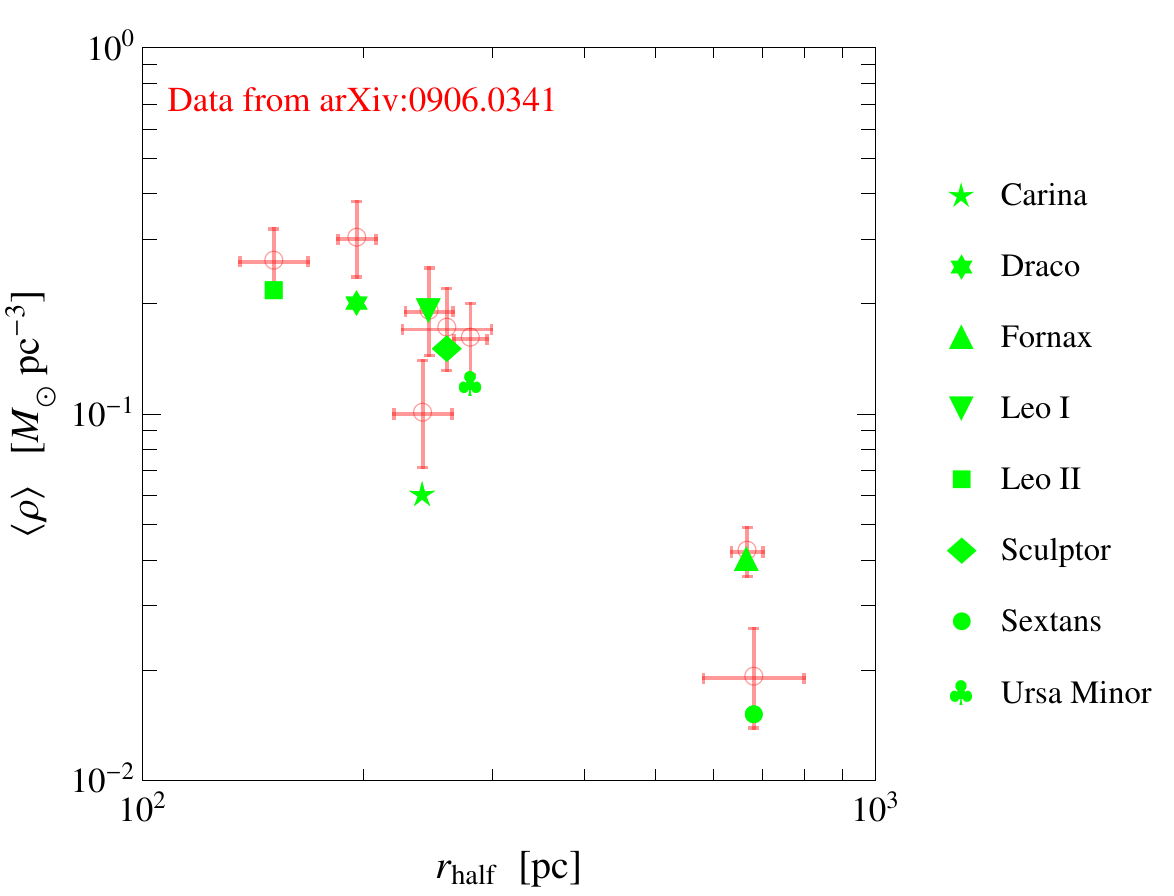}
\caption{ \textit{
Mean density within the half-light radius, $\langle \rho\rangle = 3M(r_{\rm half})/4\pi r_{\rm half}^3$. 
We compare the values obtained in our model for $m_{f} = 200$ eV with those extracted in ref.~\cite{Walker:2009zp} from 
a numerical Markov chain Monte Carlo analysis.
}}
\label{fig:HalfDensity}
\end{figure}
{Finally, in fig.~\ref{fig:HalfDensity} we analyze the mean density within the half-light radius, $\langle \rho\rangle = 3M(r_{\rm half})/4\pi r_{\rm half}^3$. In particular, we compare the values obtained in our model for $m_{f} = 200$ eV with the values extracted in ref.~\cite{Walker:2009zp} from a numerical Markov chain Monte Carlo method. Remarkably, we report a very good agreement for all the eight classical dwarf spheroidal galaxies.}

\section{Statistical mechanics of self-gravitating fermions}\label{app:FermiDirac}

{The assumption underlying this paper is that in the present Universe a galactic structure  
can be described  by a statistical equilibrium state of a self-gravitating Fermi gas of DM particles.
In this appendix we review the corresponding  statistical description. For a more detailed analysis, see refs.~\cite{LyndenBell:1966bi,Chavanis:2002rj,Chavanis:2002yv,Chavanis:2002D,ChavanisBho}. This appendix is meant to be a quick summary of some of the results obtained in these papers. 

The number density in phase-space, i.e.\ the number of DM particles per unit phase-space volume at a given conformal time $\tau$ is\footnote{It is customary to use comoving coordinates $\vec{x} = \vec{r}/a(t)$ and conformal time $d\tau = dt/a(t)$, where $a(t)$ is the scale factor. As a consequence, the comoving velocity is
$\vec{v}\equiv d\vec{x}/d\tau$, and the canonical momentum conjugate with $\vec{x}$ is $\vec{p}=a(t)m_f \vec{v}$.}
\begin{equation}
\frac{dN}{d\vec{x}d\vec{p}} = \frac{g}{h^3}f(\vec{x},\vec{p},\tau)~,
\end{equation}
where $g=2s +1$ is the number of internal (spin) degrees of freedom. In this paper we take $g=2$. 
The function $f(\vec{x},\vec{p},\tau)$ represents, by definition, the phase-space density. 
For a given galaxy, it is reasonable to assume that at the present time the phase-space density reaches a time-independent equilibrium form.
The dynamics leading to this equilibrium configuration is governed by the Liouville theorem which asserts that 
for a dissipationless and collisionless system the phase-space density is constant, $df/d\tau=0$. Writing explicitly the derivatives, one obtains the Vlasov equation
\begin{equation}
\frac{df}{d\tau} = \frac{\partial f}{\partial\tau} 
+ \frac{\vec{p}}{am_f}\cdot \frac{\partial f}{\partial\vec{x}}  -a m_f \vec{\nabla}\Phi\cdot  \frac{\partial f}{\partial\vec{p}} = 0~,
\end{equation}
where we used $d\vec{p}/d\tau = -a m_f \vec{\nabla}\Phi$, with $\Phi$ the gravitational potential. 
Starting from an arbitrary initial condition far from equilibrium, the Vlasov equation 
develops a complicated mixing process in phase-space, known as phase-space mixing. 
This process begins at matter-radiation equality, when density perturbations 
start growing shaped  by the action of gravity. The only practical way of integrating the Vlasov
equation is by N-body simulations; 
however, we are interested 
in a probabilistic description, i.e.\ we want to know the most probable 
distribution of self-gravitating fermions at statistical equilibrium. 
To determine the equilibrium distribution of the system,
it is possible 
to introduce an entropy functional like in ordinary statistical mechanics. The statistical equilibrium state 
is obtained by maximizing the entropy. 
As shown in  refs.~\cite{LyndenBell:1966bi,Chavanis:2002rj,Chavanis:2002yv,Chavanis:2002D}, critical points of the entropy 
correspond to the Fermi-Dirac distribution. Therefore, regardless of the dynamics of the system, we can directly focus on the Fermi-Dirac distribution since we are only interested in the final equilibrium state.\footnote{It is interesting to notice that the Fermi-Dirac distribution corresponds only to a
 critical point of the entropy. In order to establish if a given configuration correctly describes a stationary equilibrium state, a further analysis of the second variation of the entropy is mandatory. This is unnecessary in the degenerate Fermi limit, since it is possible to show that in this case the entropy is always maximized (the degenerate limit corresponds to the ground state of the system). However, this observation may be relevant for the case of larger galaxies under the assumption that they correspond to non-degenerate equilibrium configurations.}
 In the following, we take a closer look the the Fermi-Dirac distribution describing a system of self-gravitating fermions.
  Even if in this paper we are mainly interested in the degenerate limit, we discuss the general situation 
  at a given non-zero temperature $T$. The aim of this approach is to properly define the degeneracy temperature used in section~\ref{sec:Discussion}.}

\subsection{Fermi-Dirac distribution at temperature $T$}\label{app:GeneralFD}

The non-relativistic Fermi-Dirac distribution 
describing a system of self-gravitating fermions is given by
\begin{equation}\label{eq:FDStatistic}
f_{\rm FD} = \frac{1}{
1+
\exp
\left[
\left(
\frac{p^2}{2m_f} + m_f\Phi -\mu
\right)/k_{\rm B}T
\right]
}~,
\end{equation}
where $\Phi$ is the gravitational potential and $\mu$ the chemical potential. In order to provide a closed description of the system,
we need to couple eq.~(\ref{eq:FDStatistic}) with the Poisson equation $\triangle \Phi = 4\pi G \rho$. To achieve this goal, we need an explicit expression 
for the mass density $\rho$;
this is given by $\rho = nm_f$, where $n$ is the number density of the system 
\begin{equation}
n = 2\int_0^{\infty}
\frac{1}{
1+
\exp
\left[
\left(
\frac{p^2}{2m_f} + m\Phi -\mu
\right)/k_{\rm B}T
\right]
}\frac{4\pi p^2}{h^3}dp~.
\end{equation}
Introducing the Fermi integral $\mathcal{I}_n(t)\equiv \int_0^{\infty}x^n/(1+te^x)dx$ we obtain
\begin{equation}\label{eq:generalRHO}
\rho = 
\frac{8\pi \sqrt{2}m_f^{5/2}(k_{\rm B}T)^{3/2}}{h^3}\mathcal{I}_{1/2}(\lambda e^{m_f\Phi/k_{\rm B}T})~,
\end{equation}
where $\lambda \equiv e^{-\mu/k_{\rm B}T}$. The Poisson equation, in spherical coordinates, is
\begin{equation}\label{eq:PoissonFD}
\frac{1}{r^2}\frac{d}{dr}\left(r^2\frac{d\Phi}{dr}\right) =
\frac{32\pi^2 G \sqrt{2}m_f^{5/2}}{h^3 \,(k_{\rm B}T)^{-3/2}}\,\mathcal{I}_{1/2}(\lambda e^{{m_f\Phi}/{k_{\rm B}T}})~.
\end{equation}
Eq.~(\ref{eq:PoissonFD}) can be rewritten in a more convenient form 
by rescaling the radius according to 
\begin{equation}\label{eq:rescalingradius}
\xi = \left[
\frac{(4\pi)^22\sqrt{2}Gm_f^{7/2}(k_{\rm B}T)^{1/2}}{h^3}
\right]^{1/2}r~,
\end{equation}
and introducing the variable $\psi \equiv m_f(\Phi - \Phi_0)/k_{\rm B}T$, with $\Phi(0)\equiv \Phi_0$. We find
\begin{equation}\label{eq:PoissonRew}
\frac{1}{\xi^2}
\frac{d}{d\xi}\left(
\xi^2\frac{d\psi}{d\xi}
\right) = \mathcal{I}_{1/2}(ke^{\psi})~,
\end{equation}
with $k\equiv \lambda e^{m_f\Phi_0/k_{\rm B}T}$ and boundary conditions $\psi(0)=\psi^{\prime}(0)=0$. The solution of eq.~(\ref{eq:PoissonRew}) -- $\psi_k(\xi)$ in the following -- depends on
the value of $k$, that controls the degree of degeneracy of the gas; as we shall see in appendix~\ref{app:TDEG}, in fact, the limit $k\to \infty$ corresponds to 
the classical limit of an isothermal gas described by the Maxwell-Boltzmann statistic, while the limit $k\to 0$ corresponds to the Fermi degenerate gas.
Notice that, using the previous definition of $k$, eq.~(\ref{eq:generalRHO}) can be rewritten as
\begin{equation}\label{eq:generalRHO2}
\rho = 
\frac{8\pi \sqrt{2}m_f^{5/2}(k_{\rm B}T)^{3/2}}{h^3}\mathcal{I}_{1/2}(ke^{\psi_k(\xi)})~.
\end{equation}
Eq.~(\ref{eq:PoissonRew}) has to be integrated from $\xi = 0$ till some value $\xi_1$ that defines, via eq.~(\ref{eq:rescalingradius}),
the total radius of the configuration $R$.\footnote{In the degenerate limit of the self-gravitating Fermi gas, it is obvious to identify 
$\xi_1$ with the first zero of the mass density $\rho(r)$ (see eq.~(\ref{eq:R}) and appendix~\ref{app:TDEG} below). However, in a generic non-degenerate configuration at a given temperature $T$, the mass density $\rho(r)$ goes to zero only asymptotically, thus requiring the introduction of an empirical cut-off. In ref.~\cite{Destri:2013pt} , for instance, the integration is performed from zero till the boundary $R_{200}$, defined as the radius where 
the mass density equals $1/200$ times the mean DM density.}  Using the Gauss's theorem at the boundary $r=R$ it is possible to obtain the condition
\begin{equation}
\left.\frac{d\Phi}{dr}\right|_{r = R} = \left.\frac{GM(r)}{r^2}\right|_{r=R}~~\Rightarrow~~
\eta \equiv \frac{GMm_f}{Rk_{\rm B}T} = \xi_1\psi^{\prime}_k(\xi_1)~,
\end{equation}
where we have introduced the normalized temperature $\eta$ and where $M=M(R)$ is the total mass of the configuration.
Using eq.~(\ref{eq:rescalingradius}), simple algebra allows to show that $\xi_1^4\eta=\mu_{\rm D}^2$, where
\begin{equation}
\mu_{\rm D} \equiv \frac{2m_f^4}{h^3}\sqrt{(4\pi)^42G^3 M R^3}~,
\end{equation}
is called \textit{degeneracy parameter}. It follows that
\begin{equation}\label{eq:ConditionRaidus}
\xi_1^5\psi_k^{\prime}(\xi_1) = \mu_{\rm D}^2~.
\end{equation}
A given configuration, with a total mass $M$ and radius $R$, is characterize by a specific value of the degeneracy parameter $\mu_{\rm D}$; the integration of eq.~(\ref{eq:PoissonRew}), therefore, can be performed till the value $\xi_1$ that satisfies the condition given by eq.~(\ref{eq:ConditionRaidus}).
 
In order to proceed in the thermodynamic description of the system, we need to compute 
the pressure and the total energy; the former, in the non-relativistic limit, is given by
\begin{equation}\label{eq:generalP}
P= \frac{8\pi 2^{3/2}m_f^{3/2}(k_{\rm B}T)^{5/2}}{3h^3}\mathcal{I}_{3/2}(ke^{\psi_k(\xi)})~,
\end{equation}
while the latter is the sum of kinetic and potential energy, $E = E_k + W$. From eq.~(\ref{eq:generalP}), using the identity
$d\mathcal{I}_n(t)/dt = -(n/t)\mathcal{I}_{n-1}(t)$ and eq.~(\ref{eq:generalRHO2}), it is straightforward to show that
\begin{eqnarray}
\frac{dP(r)}{dr}&=&-\frac{8\pi \sqrt{2}m_f^{5/2}(k_{\rm B}T)^{3/2}}{h^3}\mathcal{I}_{1/2}(ke^{\psi_k(\xi)})\frac{d\Phi}{dr}\nonumber \\
&=& -\rho(r)\frac{d\Phi}{dr}~,
\end{eqnarray}
that is nothing but the condition for hydrostatic equilibrium.

The kinetic energy is given by the following integral $E_k = (3/2)\int P d\vec{x}$, from which we get
\begin{equation}
\frac{RE_k}{GM^2} = \frac{\xi_1^7}{\mu_{\rm D}^4}\int_0^{\xi_1}\mathcal{I}_{3/2}(ke^{\psi_k(\xi)})\xi^2 d\xi~.
\end{equation}
The potential energy can be obtained from the virial theorem $2E_k + W = 3VP(R)$, with $V = 4\pi R^3/3$ and $P(R)$ given by eq.~(\ref{eq:generalP})
with $\xi = \xi_1$. We find 
\begin{equation}
\frac{RW}{GM^2} = \frac{2\xi_1^{10}}{3\mu_{\rm D}^4}\mathcal{I}_{3/2}(ke^{\psi_k(\xi_1)})-\frac{2RE_k}{GM^2}~.
\end{equation}
The normalized energy is therefore given by 
\begin{eqnarray}
-\frac{RE}{GM^2} &=&
\frac{\xi_1^7}{\mu_{\rm D}^4}\int_0^{\xi_1}\mathcal{I}_{3/2}(ke^{\psi_k(\xi)})\xi^2 d\xi \nonumber \\ &-& 
\frac{2\xi_1^{10}}{3\mu_{\rm D}^4}\mathcal{I}_{3/2}(ke^{\psi_k(\xi_1)})~.
\end{eqnarray}

\subsection{The degeneracy temperature $T_{\rm DEG}$}\label{app:TDEG}

\subsubsection{Limit of a classical isothermal gas}\label{app:Iso}

The Fermi-Dirac distribution in eq.~(\ref{eq:FDStatistic})
\begin{equation}
f = \frac{1}{1+ k\exp\left[\frac{p^2}{2m_fk_{\rm B}T} + \psi(\xi)\right]}
\end{equation}
 reduces, in the limit $k\to \infty$, to a Maxwell-Boltzmann distribution $f= (1/k)e^{-[p^2/2m_fk_{\rm B}T + \psi(\xi)]}$.
 In this limit, using the approximation $\mathcal{I}_n(t) \approx (1/t)\Gamma(n+1)$, the mass density in eq.~(\ref{eq:generalRHO2}) takes the form
 \begin{equation}\label{eq:DensityIsothermal}
 \rho = \frac{2(2\pi)^{3/2}m_f^{5/2}(k_{\rm B}T)^{3/2}}{
 h^3k
 }e^{-\psi(\xi)}
 \equiv \rho_0 e^{-\psi(\xi)}~,
 \end{equation}
and, after simple algebra, the Fermi-Dirac distribution can be recast as follows
\begin{equation}
f = \frac{h^3}{2(2\pi)^{3/2}m_f^{5/2}(k_{\rm B}T)^{3/2}}\rho_0 
e^{-\psi(\xi)}e^{-p^2/2m_fk_{\rm B}T}~.
\end{equation}
The pressure in eq.~(\ref{eq:generalP}) reduces to
\begin{equation}
P=\frac{2(2\pi)^{3/2}m_f^{5/2}(k_{\rm B}T)^{3/2}}{
 h^3k
 }e^{-\psi(\xi)}\frac{k_{\rm B}T}{m_f} = \frac{k_{\rm B}T}{m_f}\rho~,
\end{equation}
that is the equation of state of a self-gravitating isothermal gas. Finally, the Poisson 
equation~(\ref{eq:PoissonRew}) takes the form\footnote{See also ref.~\cite{deVega:1996kr} for a derivation of this equation using a path integral approach.}
\begin{equation}\label{eq:PoissonBoltzmann}
\frac{1}{\xi^2}\frac{d}{d\xi}
\left(
\xi^2 \frac{d\psi}{d\xi}
\right) = e^{-\psi}
\end{equation}
where $\xi = (4\pi Gm_f\rho_0/k_{\rm B}T)^{1/2}r$ is the the rescaled radius. Eq.~(\ref{eq:PoissonBoltzmann}) is a Boltzmann-Poisson equation that can be solved numerically.

From eq.~(\ref{eq:DensityIsothermal}) it follows 
\begin{equation}\label{eq:DensityRev}
\frac{1}{k}
=
\left[
\frac{h^2}{2\pi m_f k_{\rm B}}\left(\frac{\rho_0}{2m_f}\right)^{2/3}
\right]^{3/2}\frac{1}{T^{3/2}}~;
\end{equation}
therefore, if we define the degeneracy temperature 
\begin{equation}
T_{\rm DEG} = \frac{h^2}{2\pi m_f k_{\rm B}}\left(\frac{\rho_0}{2m_f}\right)^{2/3}~,
\end{equation}
eq.~(\ref{eq:DensityRev}) becomes $1/k = (T_{\rm DEG}/T)^{3/2}$, and the Maxwell-Boltzmann limit $k\to \infty$ reduces 
to the condition $T\gg T_{\rm DEG}$.

\subsubsection{The completely degenerate limit}

This is the limit where $k\to 0$. Using the approximation 
$\mathcal{I}_n(t)\approx (-\log t)^{n+1}/(n+1)$,  it is straightforward to obtain 
from the general description outlined 
in appendix~\ref{app:GeneralFD} all the formulas used in section~\ref{sec:FermiGasQ}.

\end{document}